\documentclass[10pt, prd,twocolumn, nofootinbib,preprint,superscriptaddress]{revtex4}
\pdfoutput=1
\usepackage{amsmath,amssymb}
\usepackage{epsfig}
\usepackage{graphicx}
\usepackage[usenames,dvipsnames]{color}
\usepackage{subfigure}
\usepackage{slashed}
\usepackage[colorlinks,citecolor=blue]{hyperref}
\usepackage{color}

\begin{document}
	\title{Singlet-Doublet Self-interacting Dark Matter and Radiative Neutrino Mass}
	
	%%%%%%%%%   Authors   %%%%%%%%%%%%
	\author{Debasish Borah}
	\email{dborah@iitg.ac.in}
	\affiliation{Department of Physics, Indian Institute of Technology Guwahati, Assam 781039, India}
	\author{Manoranjan Dutta}
	\email{ph18resch11007@iith.ac.in}
	\affiliation{Department of Physics, Indian Institute of Technology Hyderabad, Kandi, Sangareddy 502285, Telangana, India}
	\author{Satyabrata Mahapatra}
	\email{ph18resch11001@iith.ac.in}
	\affiliation{Department of Physics, Indian Institute of Technology Hyderabad, Kandi, Sangareddy 502285, Telangana, India}
	\author{Narendra Sahu}
	\email{nsahu@phy.iith.ac.in}
	\affiliation{Department of Physics, Indian Institute of Technology Hyderabad, Kandi, Sangareddy 502285, Telangana, India}
	
	\begin{abstract}
		Self-interacting dark matter (SIDM) with a light mediator is a promising scenario to alleviate the small-scale problems of the cold dark matter paradigm while being consistent with the latter at large scales, as suggested by astrophysical observations. This, however, leads to an under-abundant SIDM relic due to large annihilation rates into mediator particles, often requiring an extension of the simplest thermal or non-thermal relic generation mechanism. In this work, we consider a singlet-doublet fermion dark matter scenario where the singlet fermion with a light scalar mediator gives rise to the velocity-dependent dark matter self-interaction through a Yukawa type attractive potential. The doublet fermion, by virtue of its tiny mixing with the singlet, can be long-lived and can provide a non-thermal contribution to the singlet relic at late epochs, filling the deficit in the thermal relic of singlet SIDM. The light scalar mediator, due to its mixing with the standard model Higgs, paves the path for detecting such SIDM at terrestrial laboratories leading to constraints on model parameters from CRESST-III and XENON1T experiments. Enlarging the dark sector particles by two more singlet fermions and one scalar doublet, all odd under an unbroken $\mathcal{Z}_2$ symmetry can also explain non-zero neutrino mass in scotogenic fashion.
	\end{abstract}	
	\maketitle
	%\flushbottom
	
	\section{Introduction}
	\label{intro}
	There is ample evidence suggesting the existence of a non-luminous, non-baryonic form of matter in the present Universe, known as the Dark Matter (DM). Satellite-borne experiments like Planck and WMAP, which have measured anisotropies in the cosmic microwave background (CMB) very precisely, predict that DM makes up around one-fourth ($26.8\%$) of the present energy density of the Universe. DM abundance is conventionally expressed as~\cite{Aghanim:2018eyx}: $\Omega_{\text{DM}} h^2 = 0.120\pm 0.001$ 
	at 68\% CL, where $\Omega_{\rm DM}$ is the DM density parameter and $h = \text{Hubble Parameter}/(100 \;\text{km} ~\text{s}^{-1} 
	\text{Mpc}^{-1})$ is the reduced Hubble parameter. Similar evidences for DM exist in astrophysical observations at galactic and cluster scales as well~\cite{Zwicky:1933gu, Rubin:1970zza, Clowe:2006eq}. Note that the estimate of DM abundance by Planck is based on the standard model of cosmology or the ${\rm \Lambda CDM}$ model, which has been very successful in describing the Universe at large scales $(\geq \mathcal{O}(\rm Mpc))$. In $\Lambda{\rm CDM}$ model, $\Lambda$ denotes the cosmological constant or the dark energy component of the Universe, and CDM refers to cold dark matter. Dark matter in $\Lambda{\rm CDM}$ model is assumed to be a 
	pressure-less, collision-less fluid that provides the required gravitational potential for structure formation in the early Universe. As no standard model (SM) particle mimics the properties that a DM candidate is expected to have, 
	several beyond standard model (BSM) scenarios have been proposed to explain DM. Amongst different possibilities, the weakly interacting massive particle (WIMP) 
	paradigm has been very popular. A WIMP candidate has mass and interactions in the typical electroweak regime and naturally satisfies the correct DM relic abundance from its thermal freeze-out. This remarkable coincidence is often referred to 
	as the \textit{WIMP Miracle}~\cite{Kolb:1990vq}.
	
	While ${\rm \Lambda CDM}$ predictions are in remarkable agreement with large scale structures of the Universe, at small scales, it faces severe challenges from several observations, leading to the core-cusp problem, the problem of missing satellites and the too-big-to-fail problem etc. For recent reviews on these issues and possible solutions, see~\cite{Tulin:2017ara, Bullock:2017xww}. One 
	interesting solution to these small-scale anomalies was proposed by Spergel and Steinhardt \cite{Spergel:1999mh} where they considered 
	self-interacting dark matter (SIDM) as an alternative to conventional collision-less CDM of $\Lambda{\rm CDM}$; see~\cite{Carlson:1992fn, deLaix:1995vi} 
	for earlier studies. The self interacting nature of DM is often quantified through the ratio of self-interacting cross-section of SIDM to 
	its mass as $\sigma/m \sim 1 \; {\rm cm}^2/{\rm g} \approx 2 \times 10^{-24} \; {\rm cm}^2/{\rm GeV}$ \cite{Buckley:2009in, Feng:2009hw, Feng:2009mn, 
		Loeb:2010gj, Zavala:2012us, Vogelsberger:2012ku}. The advantage of SIDM is that it resolves the small-scale anomalies of the $\Lambda{\rm CDM}$ model, 
	yet matches with the CDM halos at large radii consistent with observations. This is because of the significant dependency of self-interacting 
	cross-section on DM velocity. Studies of SIDM simulations suggest that self-interaction is stronger at smaller DM velocities and 
	thus have a large impact on small scale structures while it is in agreement with CDM predictions at larger scales with larger 
	DM velocities\cite{Buckley:2009in, Feng:2009hw, Feng:2009mn, Loeb:2010gj, Bringmann:2016din, Kaplinghat:2015aga, Aarssen:2012fx, Tulin:2013teo}. 
	From a particle physics point of view, such a velocity-dependent self-interaction can be naturally realised in models of DM with a light BSM mediator. If there exists some coupling of this light mediator with SM particles, it ensures thermal equilibrium between the dark sector and the SM bath in the early Universe. The same coupling can also be probed at direct search experiments as well~\cite{Kaplinghat:2013yxa, DelNobile:2015uua}. 
	Several model building efforts have been made to realise such scenarios. For example, see~\cite{Kouvaris:2014uoa, Bernal:2015ova, Kainulainen:2015sva, 
		Hambye:2019tjt, Cirelli:2016rnw, Kahlhoefer:2017umn,Dutta:2021wbn,Borah:2021yek,Borah:2021pet} and references therein. 
	
	Another shortcoming of the SM of particle physics is that it also fails to explain the origin of neutrino mass and mixing, which has been established by the neutrino oscillation experiments \cite{Zyla:2020zbs, Mohapatra:2005wg}. Within the SM, there is no way for the left-handed neutrinos to couple with the SM Higgs field through a renormalisable operator, as there is no right-handed neutrino field. Therefore, BSM frameworks must be invoked to explain the neutrino mass. 
	%Conventional neutrino mass models based on popular seesaw mechanisms \cite{Minkowski:1977sc, GellMann:1980vs, Mohapatra:1979ia, Schechter:1980gr, Mohapatra:1980yp, Lazarides:1980nt, Wetterich:1981bx, Schechter:1981cv, Foot:1988aq} typically involve introduction of heavy fields.
	%{\bf Remove refs. 33-41 from the draft. These refs. are not required for our purpose. }
	
	In this paper, in order to accommodate a viable SIDM component of the Universe along with an explanation of tiny neutrino mass, we consider a fermionic 
	singlet-doublet extension of the SM. Singlet-doublet fermion DM models have been well studied in the typical WIMP paradigm~\cite{Freitas:2015hsa,Cynolter:2015sua, Calibbi:2015nha, Abe:2014gua, Cheung:2013dua, Cohen:2011ec, Enberg:2007rp, DEramo:2007anh, Banerjee:2016hsk, DuttaBanik:2018emv, Horiuchi:2016tqw, Restrepo:2015ura, Badziak:2017the,Betancur:2020fdl,Abe:2019wku, Abe:2017glm,Borah:2021khc,Bhattacharya:2017sml,Barman:2019tuo,Bhattacharya:2018cgx,Bhattacharya:2016rqj,Bhattacharya:2015qpa,
		Bhattacharya:2018fus,Calibbi:2018fqf, Konar:2020wvl,Konar:2020vuu, Dutta:2020xwn, Dutta:2021uxd}, where the mixing between the singlet and the doublet fermion fields are sizeable.
	%{\bf Remove refs. 42 and 51.}
	On the contrary, 
	here we consider an extremely small singlet-doublet mixing so that the DM is dominantly composed of the singlet. The self-interaction among the DM is achieved by introducing an additional light scalar field which gives rise to an attractive Yukawa-type potential. Moreover, the light scalar mixes with the SM Higgs and paves a way to detect DM at terrestrial laboratories. The non-zero coupling of the dark matter with the visible sector brings the dark sector to thermal equilibrium. However, the requirement of large self-interaction among DM makes the thermal relic of the singlet to be under-abundant. The small mixing between the singlet and doublet fermions makes the latter long-lived, allowing the doublet to decay back to the singlet DM after the singlet component freezes out, thus fulfilling the relic deficit. We then show that the sub-eV neutrino mass can be generated in a scotogenic setup \cite{Ma:2006km} with the help of SIDM and its heavier counterparts and an additional BSM scalar doublet, all odd under an in-built $\mathcal{Z}_2$ symmetry.
	
The paper is organised as follows. In section \ref{sec2}, we discuss the model for singlet-doublet dark matter followed by a discussion on the scotogenic setup for neutrino mass generation in section \ref{sec_nu}. This is followed by a discussion on charged lepton flavour violation in section \ref{lfv}. The production and relic density of dark matter is discussed in section \ref{sec3} followed by discussion on the dark matter self-interaction  in\ref{sec4}. Then we discuss the direct and indirect detection prospects in section \ref{direct} and ~ \ref{inddet} respectively. We discuss some collider signatures of the model in section \ref{sec7} and finally conclude in section \ref{sec8}.

	%%%%%%%%%%%%%%%%%%%%%%%%%%%%%%%%%%%%%%%%%%%%%%%%%%%%%%%%%%%%%%%%%%%
	\section{Singlet-Doublet Fermion Dark Matter with Self-interactions}\label{sec2}
	%%%%%%%%%%%%%%%%%%%%%%%%%%%%%%%%%%%%%%%%%%%%%%%%%%%%%%%%%%%%%%%%%%%%
	We extend the SM by adding one vector-like fermion doublet $\Psi^T=(\psi^0, \psi^-)$ (with hypercharge 
	$Y=-1$, where we use $Q=T_3+Y/2$), and three right-handed neutrinos (RHN) $N_{R_{i}}$. A discrete $\mathcal{Z}_2$ symmetry is imposed, under which the 
	doublet $\Psi$ and all three RHNs ($N_{R_{i}}, i=1,2,3$) are odd, while all SM particles are even. All the newly added particles are also 
	singlet under $SU(3)_c$, i.e. colour neutral. To mediate velocity dependent self-interaction of DM, we add a very light scalar singlet $S$, even under the $\mathcal{Z}_2$ symmetry.
	We also add a $\mathcal{Z}_2$-odd scalar doublet $\eta$ to generate neutrino mass radiatively, which we will discuss in section~\ref{sec_nu}. The quantum numbers of 
	these BSM fields under $SU(3)_c\otimes SU(2)_L\otimes U(1)_Y \otimes \mathcal{Z}_2 $ are listed in Table \ref{tab:tab1}. 
	
	\begin{table}[h]
		%\resizebox{\linewidth}{!}{
			\begin{tabular}{|c|c|c|c|}
				\hline \multicolumn{2}{|c}{Fields}&  \multicolumn{1}{|c|}{ $\underbrace{ SU(3)_C \otimes SU(2)_L \otimes U(1)_Y}$ $\otimes   \mathcal{Z}_2 $} \\ \hline
				%\multirow{2}{*} 
				%%%%%%%%%%
				{Fermions} &  $\Psi=\left(\begin{matrix} \psi^0 \\ \psi^- \end{matrix}\right)$&  ~~1 ~~~~~~~~~~~2~~~~~~~~~~-1~~~~~~~~~ - \\ [0.5em] \cline{2-3}
				& $N_{R_{i}}$ ($i=1,2,3$) & ~~1 ~~~~~~~~~~~1~~~~~~~~~~~0~~~~~~~~~ - \\
				\hline
				\hline
				Scalars & $S=\frac{s + u + i s'}{\sqrt{2}}$  & ~~1 ~~~~~~~~~~~1~~~~~~~~~~~0~~~~~~~~~+ \\
				[0.5em] \cline{2-3}
				& $\eta = \left(\begin{matrix} \eta^+ \\ \frac{\eta^0+i \eta^{I}}{\sqrt{2}} \end{matrix}\right)$ &~1 ~~~~~~~~~~~2~~~~~~~~~~~1~~~~~~~~~~- \\
				\hline
			\end{tabular}
			%}
		\caption{Charge assignment of BSM fields under the gauge group $\mathcal{G} \equiv \mathcal{G}_{\rm SM} \otimes \mathcal{Z}_2 $,  where $\mathcal{G}_{\rm SM}\equiv SU(3)_c \otimes SU(2)_L \otimes U(1)_Y$. }
		\label{tab:tab1}
	\end{table}

	%\begin{table}[h]
	%	%\resizebox{\linewidth}{!}{
		%		\begin{tabular}{|c|c|c|c|}
			%			\hline \multicolumn{2}{|c}{Fields}&  \multicolumn{1}{|c|}{ $\underbrace{ SU(3)_C \otimes SU(2)_L \otimes U(1)_Y}$ $\otimes   \mathcal{Z}_2 $} \\ \hline
			%			%\multirow{2}{*} 
			%			%%%%%%%%%%
			%			{Fermions} &  $\Psi=\left(\begin{matrix} \psi^0 \\ \psi^- \end{matrix}\right)$&  ~~1 ~~~~~~~~~~~2~~~~~~~~~~-1~~~~~~~~~-1 \\ [0.5em] \cline{2-3}
			%			& $N _{_R}$  & ~~1 ~~~~~~~~~~~1~~~~~~~~~~0~~~~~~~~~~-1 \\
			%			\hline
			%			\hline
			%			Scalars & $H=\left(\begin{matrix} w^+ \\ \frac{h+iz}{\sqrt{2}} \end{matrix}\right)$ & ~~1 ~~~~~~~~~~~2~~~~~~~~~~1~~~~~~~~~~+\\
			%			[0.5em] \cline{2-3}
			%			& $S=\frac{s + u + i s'}{\sqrt{2}}$  & ~~1 ~~~~~~~~~~~1~~~~~~~~~~0~~~~~~~~~~+ \\
			%			[0.5em] \cline{2-3}
			%			& $\eta = \left(\begin{matrix} \eta^+ \\ \frac{\eta^0+i \eta^{I}}{\sqrt{2}} \end{matrix}\right)$ &~1 ~~~~~~~~~~~1~~~~~~~~~~0~~~~~~~~~~- \\
			%			\hline
			%		\end{tabular}
		%		%}
	%	\caption{\footnotesize{Model-II: Charge assignment of BSM fields with SM Higgs doublet under the gauge group $\mathcal{G} \equiv \mathcal{G}_{\rm SM} \otimes \mathcal{Z}_2 $  where $\mathcal{G}_{\rm SM}\equiv SU(3)_C \otimes SU(2)_L \otimes U(1)_Y$ . }}
	%	\label{tab:tab1}
	%\end{table}
	
	The Lagrangian of the model, guided by Table \ref{tab:tab1} is given by
	\begin{eqnarray}
		\label{lag}
		\mathcal{L} &=&\mathcal{L}_{\rm SM} + \overline{\Psi}~[i\gamma^{\mu}(\partial_{\mu} - i g_2 \frac{\sigma^a}{2}W_{\mu}^a - i g_1\frac{Y'}{2}B_{\mu})]~\Psi
		\nonumber \\ 
		&+&\overline{N_{R_i}}(i\gamma^\mu \partial_{\mu})N_{R_i}-M\overline{\Psi}\Psi-\frac{1}{2}M_{N_{R_i}}\overline{N_{R_{i}}}(N_{R_{i}})^c \nonumber \\
		&-& y_i \overline{\Psi}\widetilde{H}(N_{R_{i}}+(N_{R_{i}})^c ) -Y_{\alpha i} \overline{L}_\alpha \Tilde{\eta} N_{R_i} \nonumber \\ 
		&-& y_\Psi \overline{\Psi}\Psi S - y'_i \overline{N_{R_{i}}}(N_{R_{i}})^c S  + {\rm h.c.}
		+ \mathcal{L}_{\rm scalar}\,,
	\end{eqnarray}
	where $L,H$ are lepton and Higgs doublets of the SM while $\mathcal{L}_{\rm SM}, \mathcal{L}_{\rm scalar}$ are the SM Lagrangian, complete scalar Lagrangian of the model respectively. The scalar potential involving the new scalar doublet $\eta$ and the scalar singlet $S$ is given  by
	\begin{eqnarray}
		V(H,\eta , S)&=&-\mu^2_H H^\dagger H + \frac{\lambda_H}{2}(H^\dagger H)^2+\mu^2_\eta \eta^\dagger \eta + \frac{\lambda_\eta}{2}(\eta^\dagger \eta)^2\nonumber\\ \nonumber &+&\lambda_{H \eta} (H^\dagger H)(\eta^\dagger \eta)+\lambda'_{H \eta} (H^\dagger \eta)(\eta^\dagger H)\nonumber \\ 
		&+&\frac{\lambda^{''}_{H \eta}}{2}\big[(H^\dagger \eta)^2+(\eta^\dagger H)^2\big]
		 +\mu^2_{S} \left(S^\dagger S \right) \\ \nonumber &+& \lambda_S \left(S^\dagger S \right)^2 +\mu'_S S^3 + \lambda_{SH} (H^\dagger H)\left(S^\dagger S \right),
	\end{eqnarray}
	%\begin{equation}
	%\begin{aligned}
	%   V(H, \Phi_{BL}) &= -{\mu_H}^2 \left(H^\dagger H \right) + \lambda_H \left(H^\dagger H \right)^2 \nonumber \\
	%  & -{\mu_{\eta}^2 \left({\eta}^\dagger \eta \right) + \lambda_{\eta} \left({\eta}^\dagger \eta \right)^2 \nonumber  + \lambda_{H \eta} (H^\dagger H)\left({\eta}^\dagger \eta \right)\,.
		% \end{aligned}
	% \end{equation} 
where the linear terms of singlet scalar $S$ are assumed to be negligible. As shown in the Lagrangian given by Eq.~\eqref{lag}, the fermion doublet $\Psi$ being vector-like, has a bare Dirac mass M and all the three right handed neutrinos have bare Majorana mass $M_{R_i}$. We assume the Yukawa couplings $y_i$ between the doublet and singlet fermions to be very small, typically of the order $10^{-10}$. After SM Higgs and the singlet scalar $S$ acquire vacuum expectation values (VEV), the Yukawa term $y_i\overline{\Psi}\Tilde{H}N_{R_i} $ induces a Dirac mass, which is, albeit very tiny, thanks to the tiny Yukawa couplings ($y_i \sim 10^{-10}$). So, the effect of the singlet-doublet mixing is negligible and the lightest singlet fermion ($N_{R_1}$) essentially becomes the DM. The motivation for such tiny Yukawa coupling is that, it makes the doublet $\Psi$ sufficiently long-lived to produce the singlet DM at late epoch via its decay. This brings the DM relic back to the correct ballpark which is otherwise under-abundant due to strong annihilation into light scalar mediator $S$. The mixing between the singlet scalar $S$ and the SM Higgs paves a way to detect DM at terrestrial laboratories such as CRESST-III and XENON1T. As we discuss in section \ref{sec_nu}, the 
scalar doublet $\eta$ along with the singlet right handed neutrinos which are representing SIDM in our setup, give rise light neutrino masses at one loop level.

\section{Neutrino Mass} \label{sec_nu}
%%%%%%%%%%%%%%%%%%%%%%%%%%%%%%%%%%%%%%%%%%%%%%%%%%%%%%%%%%%%%%%%%%%%%
In our setup, the right-handed neutrinos ($N_{R_i}$), the lightest one of which is the SIDM, are all odd under an imposed $\mathcal{Z}_2$ symmetry. Therefore, the couplings of $N_{R_i}$ with left-handed leptons via SM Higgs are forbidden.  
%As to keep the setup minimal, SM gauge group has not been augmented by any additional gauge symmetry, the stability of the DM is guaranteed through the discrete $\mathcal{Z}_2$ symmetry. 
Consequently generation of tiny light neutrino mass at the tree level becomes unfeasible and one needs to resort to radiative scenarios. Then the simplest 
possibility is to introduce a $\mathcal{Z}_2$ odd scalar doublet $\eta$ as given in Eq.~\eqref{lag}. As a result, the light neutrino masses can be generated via 
scotogenic one-loop radiative process proposed by Ma~\cite{Ma:2006km}. 
%Scotogenic model is an extension of the SM by three copies of right handed neutrinos (RHNs) and an additional scalar doublet $\eta$, all of which are odd under an imposed $\mathcal{Z}_2$ symmetry, while the SM fields are even. In the model Lagrangian~\ref{lag}, 
The relevant terms for neutrino mass generation are identified as the bare mass terms of RHNs and the Dirac Yukawa terms that couple SM lepton doublets and RHNs via scalar doublet $\eta$, as shown in Eq. \eqref{lag}.
%\begin{equation}
%\mathcal{L} \supset -\frac{1}{2}M_{N_{R_{i}}} \overline{N_{R_{i}}} %(N_{R_{i}})^c - Y_{\alpha i} \overline{L_\alpha} \tilde{\eta} %N_{R_{i}} +  {\rm h.c.}
%\end{equation}
%The scalar potential involving the $\mathcal{Z}_2$ odd scalar doublet $\eta$ is.
%\begin{eqnarray}
%V(H,\eta)&=&-\mu^2_H H^\dagger H + \frac{\lambda_H}{2}(H^\dagger H)^2+\mu^2_\eta \eta^\dagger \eta + \frac{\lambda_\eta}{2}(\eta^\dagger \eta)^2\nonumber\\&+&\lambda_{H \eta} (H^\dagger H)(\eta^\dagger \eta)+\lambda'_{H \eta} (H^\dagger \eta)(\eta^\dagger H)+\frac{\lambda^{''}_{H \eta}}{2}\big[(H^\dagger \eta)^2+(\eta^\dagger H)^2\big]
%\end{eqnarray}
\begin{figure}[h!]
	\centering
	\includegraphics[scale=0.35]{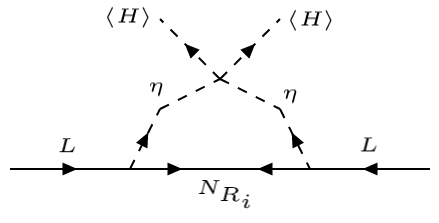}
	\caption{One-loop generation of light neutrino mass in scotogenic fashion.}
	\label{scoto}
\end{figure}
After the electroweak symmetry breaking (EWSB), the scalar doublets $H$ and $\eta$ can be parametrised as
\begin{equation}
	H \ = \ \begin{pmatrix} 0 \\  \frac{ v +h }{\sqrt 2} \end{pmatrix} , \qquad \eta \ = \ \begin{pmatrix} \eta^\pm\\  \frac{\eta^0+i\eta^I}{\sqrt 2} \end{pmatrix} \, ,
	\label{eq:idm}
\end{equation}
where $v$ is the vacuum expectation value (VEV) of SM Higgs doublet $H$, while $\eta$ does not acquire any VEV to keep the $\mathcal{Z}_2$ symmetry intact.
The neutral scalar ($\eta^0$) and pseudoscalar ($\eta^I$) acquire masses as follows.
\begin{equation}
	M^2_{R,I}=\mu^2_\eta+\frac{1}{2}(\lambda_{H \eta}+\lambda'_{H \eta}\pm \lambda''_{H \eta})~v^2,
\end{equation}
%with $+ (-)$ denoting scalar (pseudoscalar). 
Neutrino mass is induced via the one-loop diagram shown in Fig.~\ref{scoto} and is given by
\begin{widetext}
	\begin{eqnarray}
		(\mathcal{M}_\nu)_{\alpha \beta}&=&\sum_{k=1}^{3} \frac{(Y)_{i \beta} (Y)_{i \alpha}}{32 \pi^2} M_{N_{R_i}}\bigg[\frac{M^2_R}{M^2_R-M^2_{N_{R_i}}}\ln \bigg(\frac{M^2_R}{M^2_{N_{R_i}}}\bigg)-\frac{M^2_I}{M^2_I-M^2_{N_{R_i}}}\ln \bigg(\frac{M^2_I}{M^2_{N_{R_i}}}\bigg)\bigg]
	\end{eqnarray}
\end{widetext}
where $M_{N_{R_i}, i=1,2,3}$ are the mass eigenvalues of the RHN mass eigenstates $N_{R_i}, i=1,2,3$ in the internal line and the indices $\alpha, \beta 
= e, \mu, \tau$ run over the three neutrino generations. Neutrino mass vanishes in the limit of $\lambda''_{H \eta} \to 0$ as it corresponds to degenerate 
neutral scalar and pseudoscalar masses $M^2_R=M^2_I$. Thus, apart form the Yukawa couplings ($Y$) and RHN masses, the quartic coupling ($\lambda''_{H \eta}$) 
also plays a significant role in neutrino mass generation.

To include the constraints from light neutrino data in the analysis, it is often convenient to write the Yukawa couplings in the Casas-Ibarra parametrisation \cite{Casas:2001sr,Toma:2013zsa} as
\begin{equation}\label{Yukawa}
	Y = {\sqrt{\Lambda}}^{-1} R \sqrt{\hat{m_\nu}} U^\dagger_{\rm PMNS} \end{equation}
where $R$, in general, is an arbitrary complex orthogonal matrix satisfying $RR^{T}=\mathrm{I}$ that can be parametrised in terms of three complex angles ($\alpha$, $\beta$, $\gamma$). We use $R$ equal to the identity matrix ($\mathrm{I}_3$), as considering it to be complex does not alter the DM phenomenology. Here, $\hat{m_\nu} =  \textrm{Diag}(m_1,m_2,m_3)$ is the diagonal light neutrino mass matrix and the diagonal matrix $\Lambda$ is defined as $\Lambda$ = Diag ($\Lambda_1$,$\Lambda_2$,$\Lambda_3$), with
{
	%\small \begin{equation}
		%\Lambda_i=\frac{M_{N_{R_i}}}{32 \pi^2} \bigg[\frac{M^2_R}{M^2_R-M^2_{N_{R_i}}}ln \bigg(\frac{M^2_R}{M^2_{N_{R_i}}}\bigg)-\frac{M^2_I}{M^2_I-M^2_{N_{R_i}}}ln \bigg(\frac{M^2_I}{M^2_{N_{R_i}}}\bigg)\bigg].
		%\end{equation}
		
		\begin{align}
			\Lambda_i \ & = \ \frac{2\pi^2}{\lambda_5}\zeta_k\frac{2M_{N_{Ri}}}{v^2} \, , \nonumber\\
			\textrm {and}\quad \zeta_i & \ = \  \left(\frac{M_{N_{Ri}}^2}{8(M_{R}^2-M_{I}^2)}\left[L_i(M_{R}^2)-L_i(M_{I}^2) \right]\right)^{-1} \, . \label{eq:zeta}
		\end{align}
		where \begin{align}
			L_i(m^2) \ = \ \frac{m^2}{m^2-M^2_{N_{Ri}}} \: \text{ln} \frac{m^2}{M^2_{N_{Ri}}} \, .
			\label{eq:Lk}
	\end{align}}
	In Eq. (\ref{Yukawa}), 
	$U_{\rm PMNS}$ represents the usual Pontecorvo-Maki-Nakagawa-Sakata (PMNS) mixing matrix of neutrinos. 
	
	%Note that the dark matter mass directly enters into the calculation determining the dark matter mass and since DM is correlated to the doublet mass due to constraints from relic density as shown in Fig., we can get a parameter space in the plane of DM mass vs $\Psi$ mass simultaneously allowed from relic as well as neutrino mass. We show this parameter space in fig. 3. 

	\section{Lepton Flavor Violation}\label{lfv}
	In the SM, charged lepton flavour violating (CLFV) decays like $\mu \to e \gamma$ occurs at loop level and is highly suppressed by the smallness of neutrino masses and remains much beyond the current experimental sensitivity \cite{TheMEG:2016wtm}. Therefore, any future observation of such LFV decays like $\mu \rightarrow e \gamma$ will definitely be an indication of new physics beyond the SM. 
	
	\begin{figure}[h!]
		\centering
		\includegraphics[scale=0.35]{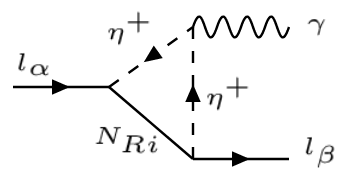}
		\caption{{Feynman diagram of CLFV decay}}
		\label{scotolfv}
	\end{figure}
	
	In the scotogenic scenario, the charged scalar doublet $\eta^+$ running in a loop along with singlet fermions can facilitate such CLFV decays, as shown in Fig. \ref{scotolfv}. This decay width of $\mu \rightarrow e \gamma$ can be calculated as\cite{Lavoura:2003xp, Toma:2013zsa},
	\begin{align}
		{\rm Br} (\mu \rightarrow e \gamma) =\frac{3 (4\pi)^3 \alpha}{4G^2_F} \lvert A_D \rvert^2 {\rm Br} (\mu \rightarrow e \nu_{\mu} \overline{\nu_e}),
	\end{align}
	where $A_D$ is given by
	\begin{equation}
		A_D = \sum_{k} \frac{(Y)^*_{ie} (Y)_{i\mu}}{16 \pi^2} \frac{1}{M^2_{\eta^+}} f (r_i),
		\label{ADMEG}
	\end{equation}
	with $r_i = M^2_{N_{R_i}}/M^2_{\eta^+}$. $f(x)$ is the loop function given by
	\begin{equation}
		f(x)=\frac{1-6x+3x^2+2x^3-6x^2\log{x}}{12 (1-x)^4}.
		\label{loop1}
	\end{equation}
	
	\begin{figure}[h!]
		\centering
		\includegraphics[width=7.5cm,height=7cm]{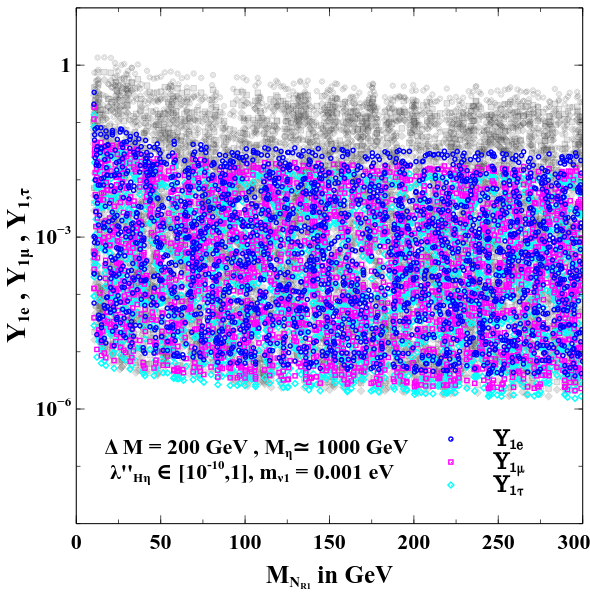}
		\caption{Yukawa couplings obtained by the Casas-Ibarra parametrisation shown against $M_{N_{R_{1}}}$. The coloured dots (blue, magenta and cyan) are allowed from all CLFV constraints while the grey coloured points which do not overlap with the colored points are ruled out.}
		\label{scotolfvyuk}
	\end{figure}
	
	The most stringent bound on such CLFV decay is from the MEG collaboration on $\text{Br}(\mu \rightarrow e \gamma) < 4.2 \times 10^{-13}$ at $90\%$ confidence level \cite{TheMEG:2016wtm}.   

Another crucial CLFV observable is the three body decay process $\mu \to 3e $. This branching fraction is given by~\cite{Toma:2013zsa}:
	\begin{eqnarray}
		\text{Br}\left(\mu \to
		e \overline{e}e\right)&=&
		\frac{3(4\pi)^2\alpha^2}{8G_F^2} ~M^2~
		 \mathrm{Br}\left(\mu \to e\nu_{\mu}
		\overline{\nu_e}\right) \,  \label{eq:l3lBR}
	\end{eqnarray}
where details of $M^2$ is given in the Appendix.~\ref{mu23e}.

Similarly, the conversion of $\mu \to e$ in nuclei might become one of the most severely constrained CLFV observables in scotogenic scenarios because of the great projected sensitivities of various collaborations.
	The conversion rate, relative to the
	the muon capture rate, can be expressed as:
	\begin{align}
		{\rm CR} (\mu \to e, {\rm Nucleus}) &= 
		\frac{p_e \, E_e \, m_\mu^3 \, G_F^2 \, \alpha_{\mathrm{em}}^3 
			\, Z_{\rm eff}^4 \, F_p^2}{8 \, \pi^2 \, Z} K^2 
		\frac{1}{\Gamma_{\rm capt}}\,.
	\end{align}
	where $Z$ and $N$ are the number of protons and neutrons in the nucleus, $Z_{\rm eff}$ is the effective atomic
	charge, $F_p$ is the nuclear matrix element and $\Gamma_{\rm capt}$ represents the total muon capture rate. Furthermore, $p_e$ and $E_e$ (taken to be $\simeq m_\mu$ in the numerical evaluation) are
	the momentum and energy of the electron and $m_\mu$ is the muon mass. The details of $K^2$ can be found in Appendix~\ref{mutoe}. 

    In Fig.~\ref{scotolfvyuk}, the $N_{R_{1}}$ Yukawa couplings obtained by the Casas-Ibarra parametrisation are shown against the mass of $N_{R_{1}}$. The coloured dots (blue, magenta and cyan) are allowed from CLFV bounds for $\mu \to e \gamma$ , $\mu \to 3e$ and ${\rm CR} (\mu \to e: {\rm Ti})$ \cite{TheMEG:2016wtm,SINDRUM:1987nra,SINDRUMII:1993gxf} , while the grey coloured points which do not overlap with the coloured points are ruled out by the CLFV constraints. For the scan, we have fixed the mass splitting between the RHNs ($\Delta M = M_{N_{R_{2,3}}}-M_{N_{R_1}}$) at $200$ GeV and $M_{\eta^+}$ is fixed around 1000 GeV. The $\lambda''_{H\eta}$ coupling, crucial for neutrino mass generation as well as determining the order of Yukawa coupling, is varied randomly in a range $[10^{-10},1]$. We consider the normal ordering and the lightest active neutrino mass is assumed to be $10^{-3}$ eV consistent with the constraint for the sum of active neutrino masses $\sum m_{\nu}= 0.12$ eV from cosmological data \cite{Aghanim:2018eyx}. From Fig. ~\ref{scotolfvyuk}, we see that the required $\overline{L} \tilde{\eta} N_{R_i}$ coupling- $Y_{\alpha i}$ roughly varies in the range $10^{-6} - 0.1$ in order to reproduce the correct light neutrino masses and mixing. We will use this information in section \ref{sec3} for calculating the relic density of 
	SIDM.  
	
	\section{Production of SIDM}\label{sec3}
\begin{figure}[htb!]
		\includegraphics[scale=0.25]{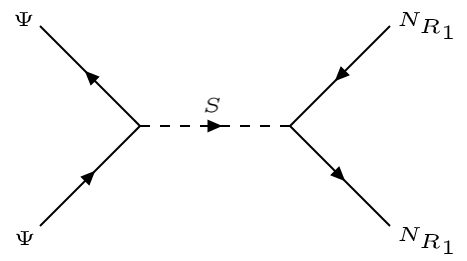}~
		\includegraphics[scale=0.3]{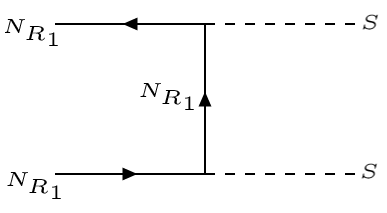}~
		\caption{Left: Feynman diagram responsible for bringing DM into thermal equilibrium, Right: Dominant annihilation mode for DM. }
		\label{fig:feyn2}
	\end{figure}

	%%%%%%%%%%%%%%%%%%%%%%%%%%%%%%%%%%%%%%%%%%%%%%%%%%%%%%%%%
	Due to electroweak gauge interactions, the vector-like fermion doublet $\Psi$ remains in thermal equilibrium at a temperature above its mass scale in the early Universe. 
	The SM gauge singlets ($N_{R_{i}}$), also come to thermal equilibrium through the process $\Psi \Psi \xrightarrow{S} 
	N_{R_{i}} N_{R_{i}}$ (shown in the left panel of Fig.~\ref{fig:feyn2}) mediated by the light scalar $S$. This is because of the large Yukawa coupling: $ y'_i 
	\sim 0.35 $ which is necessary for sufficient self interaction as discussed in section \ref{sec4}. Due to efficient annihilation of the dark matter into light 
	scalar $S$ through the process shown in the right panel of Fig.~\ref{fig:feyn2}, the thermal relic of the DM ($N_{R_{1}}$) is found to be under-abundant. 
	%The thermally averaged cross-section for this process is approximately given by:
	%\begin{equation}
	%	\langle\sigma v\rangle_{\rm DM~ DM \to~ S~S} \approx	\frac{3}{4}\frac{y'^4_1}{16\pi M^2_{\rm DM}}
	%	\label{ann}
	%\end{equation}
	The annihilation of a fermion pair through a scalar mediator (irrespective of the final states) is a p-wave process and hence velocity suppressed~\cite{Arina:2018zcq}. The thermally averaged cross-section for the most dominant annihilation process relevant to DM freeze-out is given by:
	\begin{equation}
	 \langle \sigma v \rangle_{{\rm DM ~DM} \to S~S} = \frac{3}{4}\frac{y'^4}{16\pi M^2_{DM}} v^2 \sqrt{1-\frac{M^2_S}{M^2_{{\rm DM}}} }
		\label{ann}
	\end{equation}	
	where $M_{\rm DM}= M_{N_{R_{1}}}$.
	%\begin{figure}[htb!]
	%  \includegraphics[scale=0.4]{sd.jpg}~
	% \includegraphics[scale=0.41]{psi_cs.pdf}~
	%\caption{\footnotesize{Left: Freeze-out abundance of $\Psi$, Right: Corresponding thermally averaged cross-section. }}
	%\label{self}
	%\end{figure}

	%\begin{figure}[h!]
		% \includegraphics[scale=0.45]{sv_compare.pdf}~
	%	\includegraphics[width=8cm,height=7cm]{compare.pdf}
	%	\caption{Comparison between strengths of $\langle \Psi \Psi \to {\rm SM \,SM} \rangle$ and $\langle \Psi \Psi \to S S \rangle$ as a function of mass $M_\Psi$. }
	%	\label{fig:psi_r}
	%\end{figure}
	\begin{figure}[h!]
		\includegraphics[width=8cm,height=7cm]{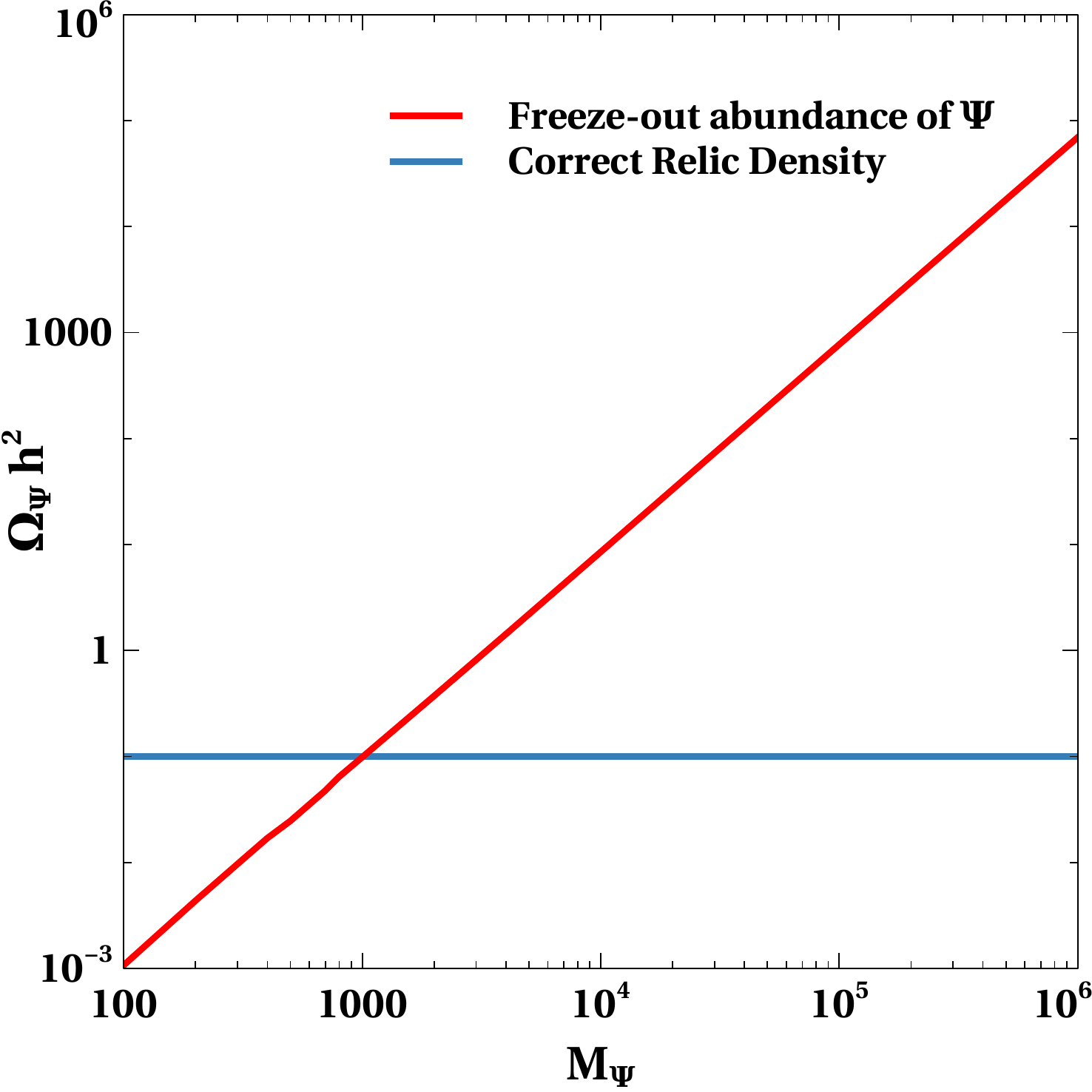}
		\caption{Freeze-out abundance of $\Psi$ as a function of its mass $M_\Psi$. }
		\label{fig:psi_relic}
	\end{figure}
	
	The thermal relic of DM $N_{R_{1}}$ is under-abundant up to DM mass $\sim 1$ TeV due to large annihilation cross-section into light mediators given by Eq.~\eqref{ann}. Thanks to the small Yukawa coupling of the Lagrangian term $y_i \Psi \Tilde{H}N_{R_{i}}$, the doublet $\Psi$ decays sufficiently late into SM Higgs and the three singlets ($N_{R_{i}}$), and since the heavier singlets $N_{R_{2,3}}$ also decay into the DM $N_{R_{1}}$ eventually, it helps in restoring the DM relic within the correct ballpark.
	%The formulas for all the relevant decay widths are given in Appendix-\ref{appendix2}.
	
	%Note that DM can also be produced from $\eta$ decay $\eta \to N_{i}l$. The decay widths of this process is constrained from charged lepton flavour violation discussed in section \ref{lfv} as the same coupling is involved in both the processes. For $\lambda''_{H\eta}=\mathcal{O}(1)$, the dominant Yukawa coupling $Y_{i\alpha}$ is of the order $\mathcal{O}$($10^{-5}$). This decay width corresponding to this coupling is calculated to be $\Gamma_{\eta \to N_{R_{i}}l} \simeq 1 \times 10^{-9}{\rm GeV}$. Due to such large decay width, the decay occurs very early, and hence the produced DM annihilates into light mediators quickly. To survive this annihilation, DM must be produced after ${\rm DM~ DM} \to S~S$ process freezes out so that there is no further depletion in DM number density and decay of the doublet $\Psi \to H N_{R_{i}}$ serves this purpose. 
	
	In the early Universe, the number density of the doublet $\Psi$ gets depleted in three ways {\it viz.} $\Psi \Psi \to {\rm S S}$, $\Psi \Psi \to N_{R_{i}} N_{R_{i}}$ and $\Psi \Psi \to {\rm SM ~SM}$. Here we assume the Yukawa coupling $y_\Psi$ of the doublet $\Psi$ with light scalar $S$ to be sufficiently small, so that $\Psi \Psi \to {\rm SM ~SM}$ always dominantly decides the freeze-out abundance of $\Psi$. 
%Even if the Yukawa coupling $y_\Psi$ were of the same order as that of $y'_1$ which responsible for sufficient self-interaction among DM, the process $\Psi \Psi \to {\rm SM ~SM}$ still dominates over $\Psi \Psi \to {\rm S ~S}$ as shown in Fig.\ref{fig:psi_r}. 
Considering $\Psi \Psi \to {\rm SM ~SM}$ to be the dominant process, the freeze-out number density of $\Psi$ can be easily calculated by implementing SM + inert fermion doublet ($\Psi$) model in \texttt{LanHEP}~\cite{SEMENOV1998124} and feeding the model files into \texttt{MicrOmegas}~\citep{Belanger:2008sj}. The freeze-out abundance of $\Psi$ as a function of mass $M_\Psi$ is shown in Fig.~\ref{fig:psi_relic}. The relevant interactions for the fermion doublet in component form are shown in Appendix-\ref{appendix1}. From Fig.~\ref{fig:psi_relic}, we see that freeze-out abundance of $\Psi$ matches the correct relic density ($\Omega_{\rm DM}h^2 =0.12$) only for mass around 1000 GeV. Since it is essentially the number density of $\Psi$ that is converted into number density of the DM at late epoch, to produce correct relic for DM $N_{R_{1}}$, the freeze-out number density of $\Psi$ must satisfy
	\begin{equation}
		\label{fact}
		\Omega_{\rm DM} h^2 = \Big(\frac{M_{\rm DM}}{M_\Psi}\Big)\Omega_\Psi h^2
	\end{equation}
Note that the inert scalar doublet $\eta$ also has Yukawa coupling to $N_{R_i}$. Depending upon the mass hierarchies of $\eta$ and $N_{R_{2,3}}$, one can decay into other, and this may affect the DM relic density because all $\mathcal{Z}_2$ odd particles will ultimately decay into DM. To obtain the relic density precisely, we need to solve the relevant coupled Boltzmann equations. We consider two different scenarios depending upon masses of $\eta$ and $N_{R_{2,3}}$.

{\bf Case-I:} $\mathbf {M_{N_{R_{2,3}}}<M_\eta < M_\Psi}$. In this case, $\eta$ can decay into all three $N_{R_i}$ with corresponding partial decay width $\Gamma_{\eta \to N_{R_i}l}$. Thermally produced $N_{R_{2,3}}$ along with those produced from decays of $\Psi$ and $\eta$ will subsequently decay into DM through the three body decay process $\Gamma(N_{R_{2,3}} \to N_{R_1}ll)$. Both these processes are simultaneously constrained from charged lepton flavour violation discussed in section \ref{lfv}. The relevant Boltzmann equation in this case is given by Eq.~\eqref{eq:BE1} below while the formulae for relevant cross-section and the decay widths are given in Appendix \ref{appendix3}. 
\begin{widetext}
	\begin{equation}
		\label{eq:BE1}
		\begin{aligned}
			\frac{dY_{N_{R_{1}}}}{dx}&=-\frac{s(M_{\rm DM})}{x^2  H(M_{\rm DM})}\big( \langle \sigma(N_{R_{1}} N_{R_{1}} \to SS) v \rangle \big) (Y^2_{N_{R_{1}}} -\big(Y^{\rm eq}_{N_{R_{1}}}\big)^2)+\frac{x}{H(M_{\rm DM})}\big( \langle \Gamma_{\Psi \rightarrow H N_{R_{1}}}\rangle Y_{\Psi} + \langle \Gamma_{\eta \rightarrow N_{R_{1}}l}\rangle Y_{\eta} \big)\\
			\frac{dY_{N_{R_{2,3}}}}{dx}&=-\frac{s(M_{\rm DM})}{x^2  H(M_{\rm DM})}\big( \langle \sigma(N_{R_{2,3}} N_{R_{2,3}} \to SS) v \rangle \big) (Y^2_{N_{R_{2,3}}} -\big(Y^{\rm eq}_{N_{R_{2,3}}}\big)^2)\\
			&+\frac{x}{H(M_{\rm DM})}\big( \langle \Gamma_{\Psi\rightarrow H N_{R_{2,3}}}\rangle Y_{\Psi} + \langle \Gamma_{\eta \rightarrow N_{R_{2,3}}l}\rangle Y_{\eta} + \langle \Gamma_{N_{R_{2,3}} \rightarrow N_{R_{1}}ll}\rangle Y_{N_{R_{2,3}}}\big)\\
			\frac{dY_\Psi}{dx}&=-\frac{s(M_{\rm DM})}{x^2  H(M_{\rm DM})} \big(\langle(\sigma(\Psi \Psi \to {\rm SM\, SM}) v\rangle+\langle\sigma(\Psi \Psi \to N_R N_R) v\rangle+\langle\sigma(\Psi \Psi \to SS) v\rangle\big) (Y^2_\Psi -\big(Y^{\rm eq}_\Psi\big)^2)\\
			&-\frac{x}{H(M_{\rm DM})}\big( \langle \Gamma_{\Psi\rightarrow H N_{R_{i}}}\rangle Y_{\Psi} \big)\\
			\frac{dY_{\eta}}{dx}&=-\frac{s(M_{\rm DM})}{x^2  H(M_{\rm DM})}\big( \langle \sigma(\eta \eta \to HH) v \rangle \big) (Y^2_{\eta} -\big(Y^{\rm eq}_{\eta}\big)^2)-\frac{x}{H(M_{\rm DM})}\big( \langle \Gamma_{\eta \rightarrow N_{R_{i}}}\rangle Y_{\eta} \big)
		\end{aligned}
	\end{equation}
\end{widetext}
In the above $x=\frac{M_{\rm DM}}{T}$, $s(M_{\rm DM})= \frac{2\pi^2}{45}g_{*s}M^3_{\rm DM}$ , $H(M_{DM})=1.67 g^{1/2}_*\frac{M^2_{DM}}{M_{\rm Pl}}$ and $<\sigma(\Psi \Psi \to {\rm SM \, SM}) v>$ represents the thermally averaged cross-section~\cite{Gondolo:1990dk} of annihilation of $\Psi$ to all SM particles, which is fixed by the SM gauge interaction of the doublet $\Psi$. Also $\langle \Gamma_{A \rightarrow B\, C\, D}\rangle$ represents the thermally averaged decay width of the process $A\rightarrow B \, C \, D$ in general. %The relevant cross-sections and the decay widths are furnished in Appendix~\ref{appendix2}.

{\bf Case-II:} $\mathbf {M_\eta < M_{N_{R_{2,3}}}<M_\Psi}$. In this case, $N_{R_{2,3}}$ can decay into $\eta$ through $\Gamma({N_{R_{2,3}} \to \eta ~l})$ and $\eta$ then decays into the DM through the process $\eta \to N_{R_1}l$, constrained by charged lepton flavour violation. The relevant Boltzmann equations in this case are given by Eq.~\eqref{eq:BE2} below. 
\begin{widetext}
	\begin{equation}
		\label{eq:BE2}
		\begin{aligned}
			\frac{dY_{N_{R_{1}}}}{dx}&=-\frac{s(M_{\rm DM})}{x^2  H(M_{\rm DM})}\big( \langle \sigma(N_{R_{1}} N_{R_{1}} \to SS) v \rangle \big) (Y^2_{N_{R_{1}}} -\big(Y^{\rm eq}_{N_{R_{1}}}\big)^2)+\frac{x}{H(M_{\rm DM})}\big( \langle \Gamma_{\Psi \rightarrow H N_{R_{1}}}\rangle Y_{\Psi} + \langle \Gamma_{\eta \rightarrow N_{R_{1}}l}\rangle Y_{\eta} \big)\\
			\frac{dY_{N_{R_{2,3}}}}{dx}&=-\frac{s(M_{\rm DM})}{x^2  H(M_{\rm DM})}\big( \langle \sigma(N_{R_{2,3}} N_{R_{2,3}} \to SS) v \rangle \big) (Y^2_{N_{R_{2,3}}} -\big(Y^{\rm eq}_{N_{R_{2,3}}}\big)^2)+\frac{x}{H(M_{\rm DM})}\big( \langle \Gamma_{\Psi\rightarrow H N_{R_{2,3}}}\rangle Y_{\Psi} - \langle \Gamma_{N_{2,3} \rightarrow \eta l}\rangle Y_{N_{R_{2,3}}}\big)\\
			\frac{dY_\Psi}{dx}&=-\frac{s(M_{\rm DM})}{x^2  H(M_{\rm DM})} \big(\langle(\sigma(\Psi \Psi \to {\rm SM \,SM}) v\rangle+\langle\sigma(\Psi \Psi \to N_R N_R) v\rangle+\langle\sigma(\Psi \Psi \to SS) v\rangle\big) (Y^2_\Psi -\big(Y^{\rm eq}_\Psi\big)^2)\\
			&-\frac{x}{H(M_{\rm DM})}\big( \langle \Gamma_{\Psi\rightarrow H N_{R_{i}}}\rangle Y_{\Psi} \big)\\
			\frac{dY_{\eta}}{dx}&=-\frac{s(M_{\rm DM})}{x^2  H(M_{\rm DM})}\big( \langle \sigma(\eta \eta \to HH) v \rangle \big) (Y^2_{\eta} -\big(Y^{\rm eq}_{\eta}\big)^2)+\frac{x}{H(M_{\rm DM})}\big( \langle \Gamma_{N_{2,3} \rightarrow \eta l}\rangle Y_{N_{R_{2,3}}} - \langle \Gamma_{\eta \rightarrow N_{R_{1}}}\rangle Y_{\eta} \big)
		\end{aligned}
	\end{equation}
\end{widetext}
%In the above $x=\frac{M_{DM}}{T}$, $s(M_{DM})= \frac{2\pi^2}{45}g_{*S}M^3_{DM}$ , $H(M_{DM})=1.67 g^{1/2}_*\frac{M^2_{DM}}{M_{Pl}}$ and $<\sigma(\Psi \Psi \to SM SM) v>$ represents the thermally averaged cross-section~\cite{Gondolo:1990dk} of annihilation of $\Psi$ to all SM particles, which is fixed by the SM gauge interaction of the doublet $\Psi$. Also  $\langle \Gamma_{\Psi\rightarrow H N_R}\rangle$ represents the thermally averaged decay width of the process $\Psi\rightarrow H N_R$. The relevant cross-sections and the decay widths are furnished in Appendix~\ref{appendix2}.

\begin{figure}[h!]
	\includegraphics[width=8cm,height=7cm]{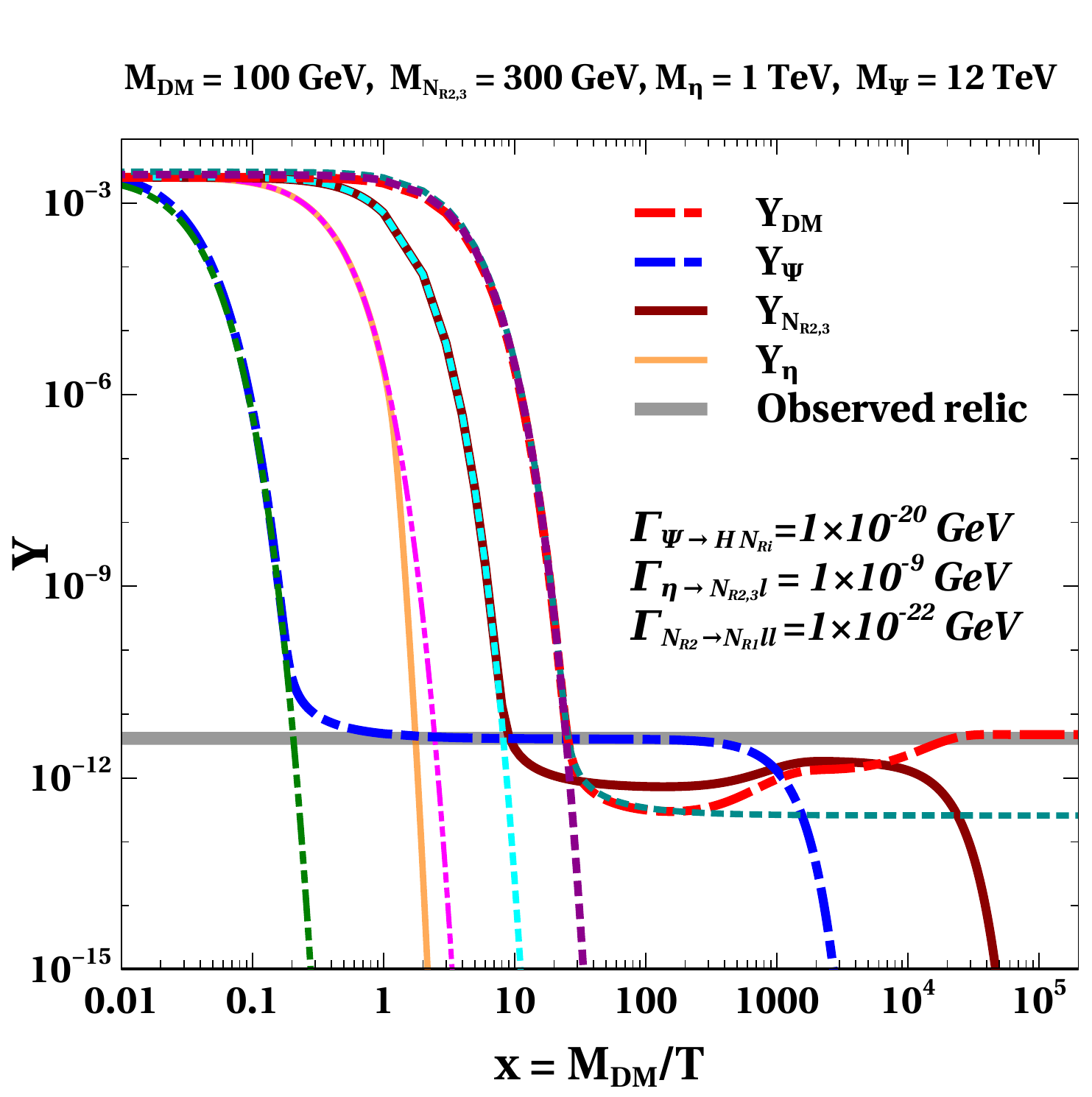}\\
	\includegraphics[width=8cm,height=7cm]{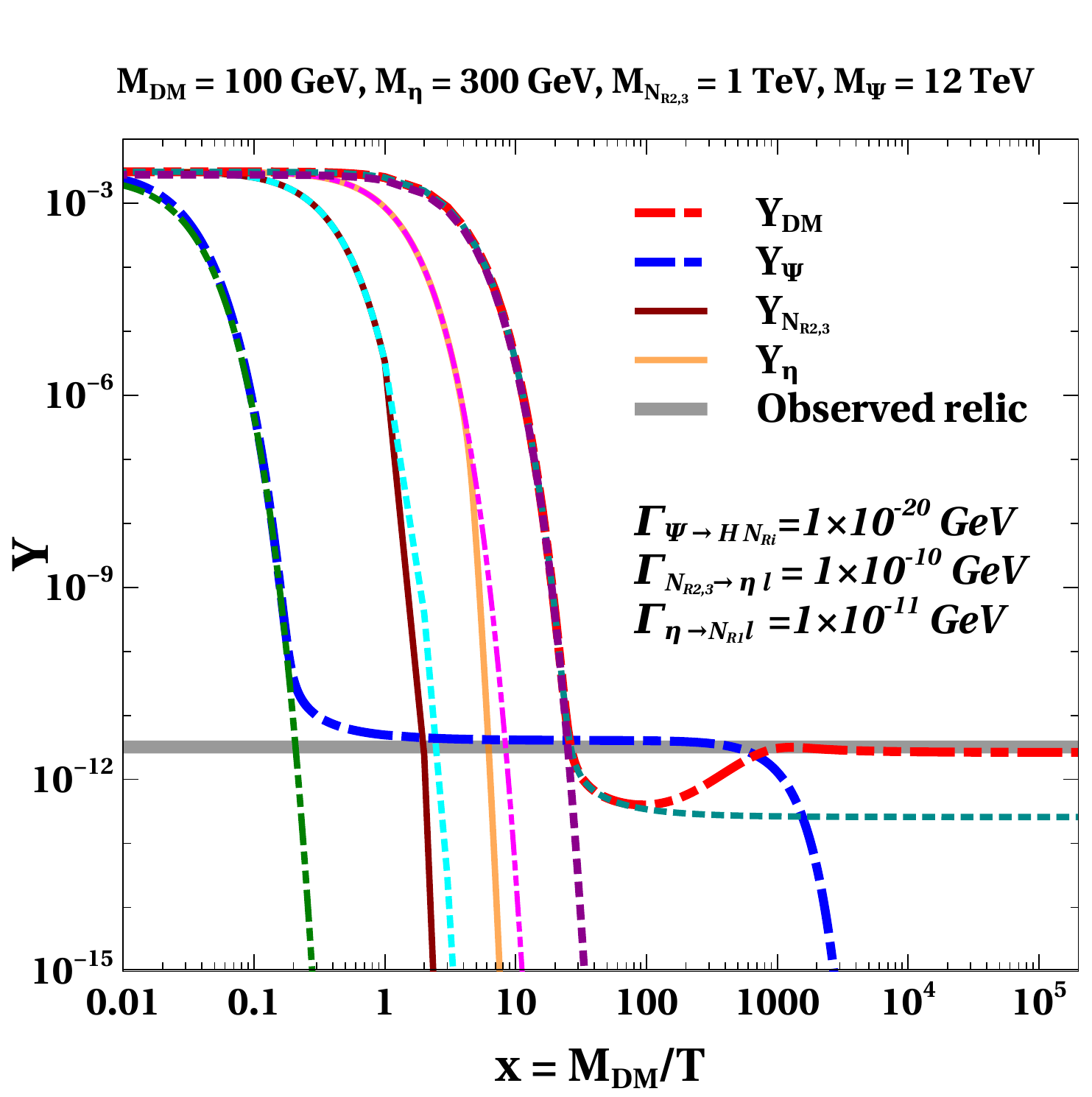}
	\caption{{Evolution of co-moving number densities of dark matter along with other dark sector particles.}}
	\label{relic2}
\end{figure}
%\begin{figure}[htb!]
% \includegraphics[scale=0.45]{sd100-12.pdf}~
%\includegraphics[scale=0.45]{sd50-30000.pdf}~
%\caption{\footnotesize{Left: Freeze-out abundance of $\Psi$, Right: Corresponding thermally averaged cross-section. }}
%\label{relic5}
%\end{figure
	
	We show the relic density evolution plots for case-I and case-II in the top and bottom panels of Fig.~\ref{relic2} for $M_{\rm DM} =$ 100 GeV, $M_\Psi$= 12 TeV, which yield correct relic density of the DM.   
	%We show the relic density by taking $\Gamma_{\eta \to N_{R_{i}}l} = 1 \times 10^{-9}{\rm GeV}$ and $\Gamma_{N_{R_{2,3}} \to N_{R{1}}ll} = 1 \times 10^{-22}{\rm GeV}$, where the parameters are consistent with light neutrino data as well as  CLFV constraints. Any DM production before DM annihilation to light mediator will leave insignificant DM relic. The decay $\Gamma_{\eta \to N_{R_{i}}l}$ is so large that it decays while still in equilibrium and does not affect the relic density significantly.
	%becomes smaller than this, $\eta$ decay becomes sufficiently late where DM annihilation to light mediator S is no longer active, and it affects the relic density significantly; hence the simple relation~\ref{fact} no longer holds good. 
	For better understanding, we show in Fig.~\ref{relic2}, the contributions from different sub-processes to the relic abundance in different colour codes as indicated in the figure inset. The evolution of number densities of $\Psi$, $\eta$, $N_{R_{2,3}}$ and the DM in light of all the processes incorporated in the Boltzmann equations given by Eq.~\eqref{eq:BE1}, Eq.~\eqref{eq:BE2} are shown in blue, orange, brown and red coloured curves respectively. Additionally, we have also shown the equilibrium distribution of $\Psi$, $\eta$, $N_{R_{2,3}}$ and the DM in green, magenta, cyan and purple coloured curves respectively and the under-abundant thermal freeze-out relic of DM is depicted by the dotted dark-cyan curve. Due to the large decay width of $\eta$, it decays while still in equilibrium ($\eta\rightarrow N_{R_{i}} l$) which is too early to mark any significant effect in the final relic of DM $N_{R_1}$. 
	
	In case-I (top panel of Fig.~\ref{relic2}), the processes $\eta \to N_{R_i}l$ and $N_{R_{2,3}} \to N_{R_1}ll$ are simultaneously constrained from charged lepton flavour violation as the same couplings are involved in all the processes. We consider $M_\eta = 1000\, {\rm GeV}$ and $M_{N_{R_{2,3}}}=300\, {\rm GeV}$ where the masses of $N_{R_{2,3}}$ are assumed to be equal for simplicity. Then, for $\lambda''_{H\eta}=\mathcal{O}(1)$ and assuming normal hierarchy of active neutrinos, the allowed Yukawa coupling $Y_{i\alpha}$ is of the order $\mathcal{O}$($10^{-5}$). The decay width corresponding to this coupling is calculated to be $\Gamma_{\eta \to N_{R_{i}}l} \simeq 1 \times 10^{-9}\,{\rm GeV}$. Due to such large decay width, $\eta$ decays while in equilibrium and DM produced at such an early epoch annihilates into light mediators quickly. Therefore, $\eta$ does not affect DM relic density. The three body decay width of two heavier singlets $N_{R_{2,3}}$ for the same parameter space considered in  $\eta \to N_{i}l$ is given by $\Gamma_{N_{R_{2,3}} \to N_{R{1}}ll} \simeq 1 \times 10^{-22}\,{\rm GeV}$. Since the decay width of $N_{R_{2,3}}$ is comparable to that of $\Gamma_{\Psi \to H N_{R_i}}$,  its effect can be seen in the plot shown in Fig.~\ref{relic2}. The decay of $\Psi$ with decay width $\Gamma_{\Psi \to H N_{R_i}} = 1\times 10^{-20}{\rm GeV}$ produces all $N_{R_{i}}$'s equally around $x\sim 2\times10^{3}$ and finally $N_{R_{2,3}}$ decays into the DM ($N_{R_{1}}$) around $x\sim 2\times 10^{4}$, producing correct final relic density for the dark matter. Note that considering inverted hierarchy of active neutrinos does not alter the consequences significantly.
	%Such small decay width of $\Psi$ can be realised with $\overline{\Psi} \tilde{H}N_{R_i}$ coupling $y_i \sim \mathcal{O}^{10^{-10}}$.It is to be noted that to get the correct relic density of DM from the decay of $\Psi$ after $\Psi$ freezes out, the decay $\Psi \to H N_R $ must occur after the process $N_R N_R \to SS$ becomes completely ineffective ($x \sim 30$ as shown in left panel of Fig.~\ref), otherwise, the produced DM from $\Psi$ decay will still undergo annihilation into light $S$ to give underabundance.
	%\begin{equation}
	In case-II (bottom panel of Fig.~\ref{relic2}), for $\lambda''_{H\eta}=\mathcal{O}(1)$ and normal hierarchy, and masses $M_\eta = 300 \,{\rm GeV}$ and $M_{N_{R_{2,3}}}=1000\, {\rm GeV}$, the decay widths are calculated to be $\Gamma_{N_{R_{2,3}} \to \eta l} \simeq 1 \times 10^{-10}\,{\rm GeV}$ and $\Gamma_{\eta \to N_{R_1}l} \simeq 1 \times 10^{-11}\, {\rm GeV}$. Due to such large decay widths, both $\eta$ and $N_{R_{2,3}}$ decays while still in equilibrium, not affecting the final DM relic at all. The correct relic of DM is entirely decided by late decay of $\Psi$ with $\Gamma_{\Psi \to H N_{R_i}} = 1\times 10^{-20}\, {\rm GeV}$.
	%\label{eq:BE1}
	%\begin{aligned}
	%\frac{dY_R}{dx}&=-\frac{s(M_\Psi)}{x^2  H(M_\Psi)}\big( \langle \sigma(N_R N_R \to SS) v \rangle \big) (Y^2_N_R -\big(Y^{eq}_N_R\big)^2)\\
	%\frac{dY_\Psi}{dx}&=-\frac{s(M_\Psi)}{x^2  H(M_\Psi)} \big(\langle(\sigma(\Psi \Psi \to SM SM) v\rangle+\langle\sigma(\Psi \Psi \to N_R N_R) v\rangle+\langle\sigma(\Psi \Psi \to SS) v\rangle\big) (Y^2_\Psi -\big(Y^{eq}_\Psi\big)^2)
	%\end{aligned}
	%\end{equation}
	
	%\newcommand{\eqdef}{\overset{\mathrm{def}}{=\joinrel=}}
	
\begin{figure}[h!]
		\includegraphics[width=8cm,height=7cm]{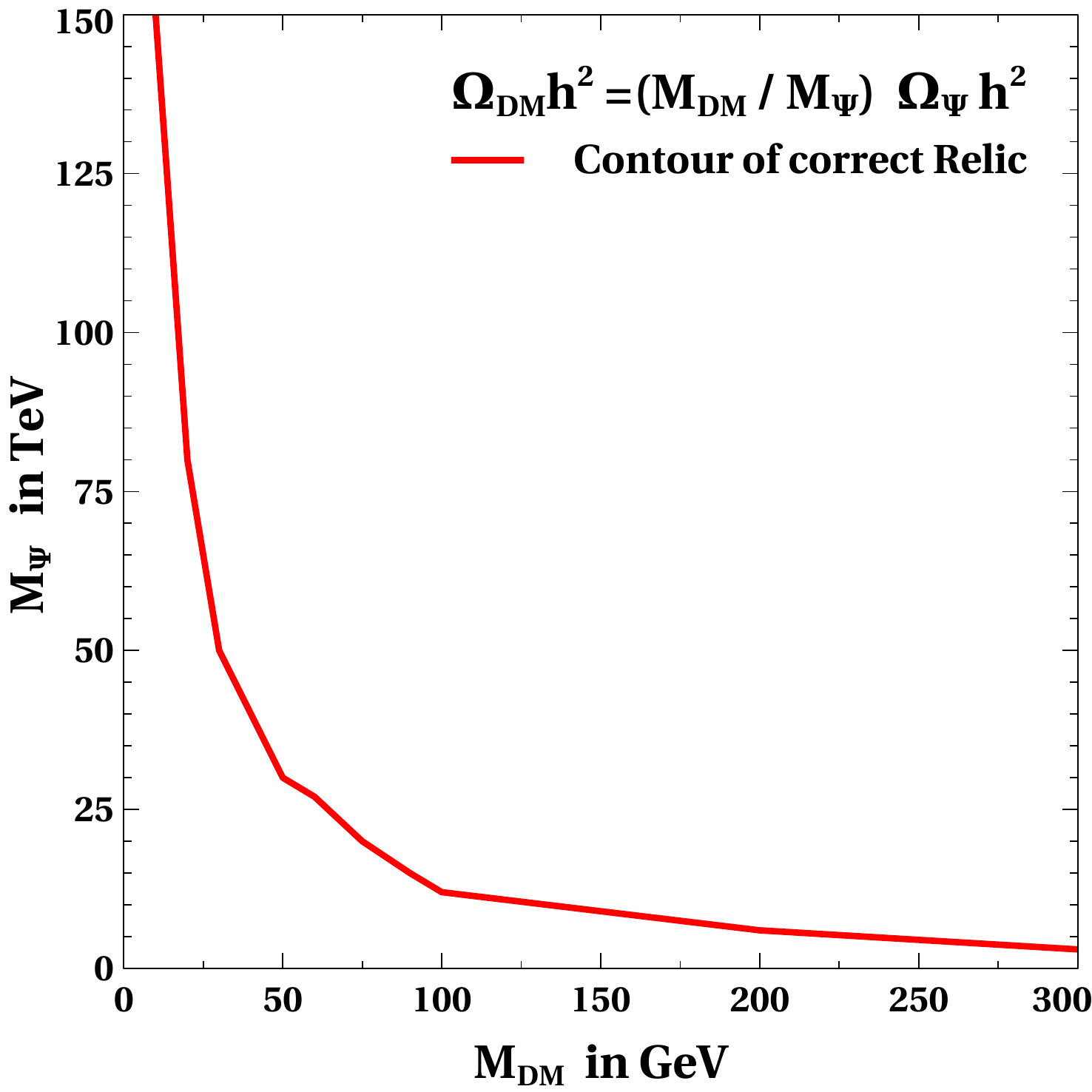}
		\caption{Contour of correct relic density for the DM ($N_{R_1}$) in the plane of $M_\Psi$ and $M_{\rm DM}$.}
		\label{relic}
	\end{figure}

	%\[\sigma (HH\rightarrow N_R N_R)  \overset{M_\Psi >> t, M_{DM}}{=\mathrel{\mkern-3mu}=} \frac{y^4}{64 \pi^2 M^2_\Psi} \] 
	As we can see from Fig.~\ref{relic2}, due to the constraints from charged lepton flavour violation, the decay processes other than $\Psi \to H N_{R_i}$ do not affect the relic significantly and the relation given by Eq.~\eqref{fact} holds for correct DM relic. Thanks to the validity of Eq.~\eqref{fact} and the fact that $\Omega_\Psi h^2$ is decided entirely by $M_\Psi$, only a certain combination of ($M_{\rm DM}, M_\Psi$) will produce the correct relic density for the DM. We show in Fig.~\ref{relic}, the contour of correct DM relic density in the plane of $M_\Psi$ versus $M_{\rm DM}$. As we can see, to obtain the correct relic density for a light DM (below 10 GeV), $M_\Psi$ must be very heavy (above 150 TeV) as expected from Eq.~\eqref{fact}. As $M_{\rm DM}$ increases, $M_\psi$ decreases satisfying Eq.~\eqref{fact} and for $M_{\rm DM} =$300 GeV, the corresponding $M_\psi =$ 3 TeV. For $M_{\rm DM}\sim$ 1 TeV, the relic density in the correct ballpark can directly be satisfied from its thermal freeze-out, beyond which thermal DM relic is over-abundant. 
	%However, $M_{\rm DM} >$ 300 GeV region is disfavoured from both self-interaction and direct search which we will discuss in section~\ref{sec4}and section \ref{sec5} respectively. 
	%%%%%%%%%%%%%%%%%%%%%%%%%%%%%%%%%%%%%%%%%%%%%%%%%%%%%%%%%%%%%%%%%%%%%%%%%%%%%%%%%%%%%%%%%%%
	%%%%%%%%%%%%%%%%%%%%%%%%%%%%%%%%%%%%%%%%%%%%%%%%%%%%%%%%%%%%%%%%%%%%%%%%%%%%%%%%%%%%%%%%%%%%
	
	\section{Dark Matter Self-interaction}\label{sec4}
\begin{figure}[h!]
		\centering
		\includegraphics[width = 55mm]{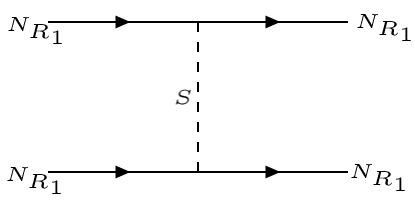}
		\caption{Feynman diagram for elastic DM self-interaction.}
		\label{sidm}
	\end{figure}

\begin{figure}[h!]
		\includegraphics[width=8cm,height=8cm]{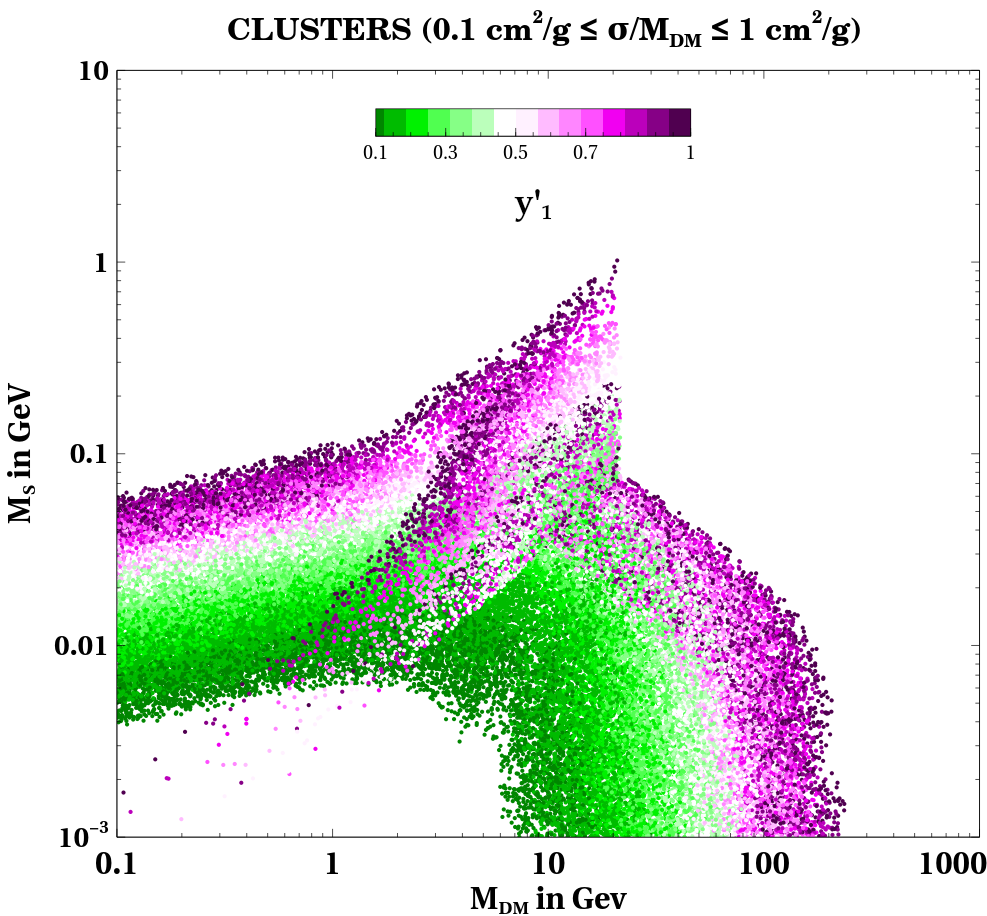}
		\hfil
		\includegraphics[width=8cm,height=8cm]{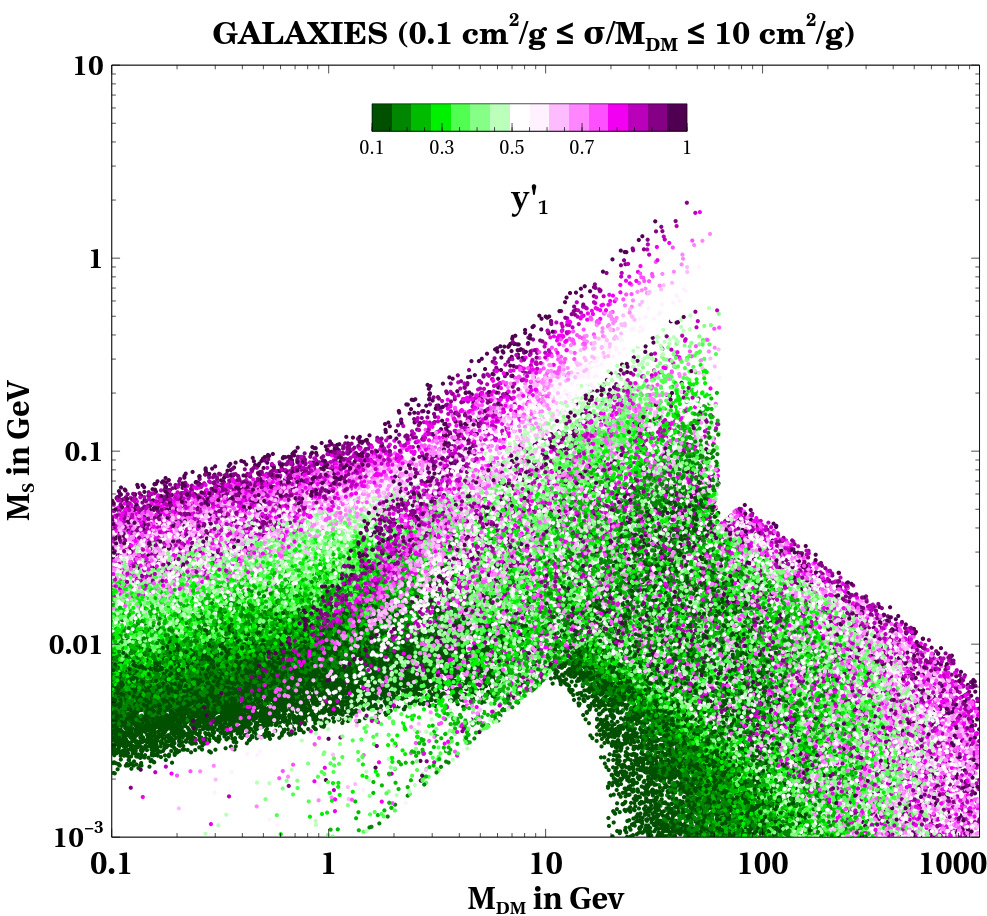}
		\caption{Top panel: Cluster;  Bottom panel: Galaxy;  Self-interaction cross-section in the range $0.1-1 \; {\rm cm}^2/{\rm g}$ for clusters ($v\sim1000 \; {\rm km/s}$) and $0.1-10 \; {\rm cm}^2/{\rm g}$ for galaxies ($v \sim 200 $ km/s.)}
		\label{sidm1}
	\end{figure}

\begin{figure}[h!]	
		\includegraphics[width=8cm,height=8cm]{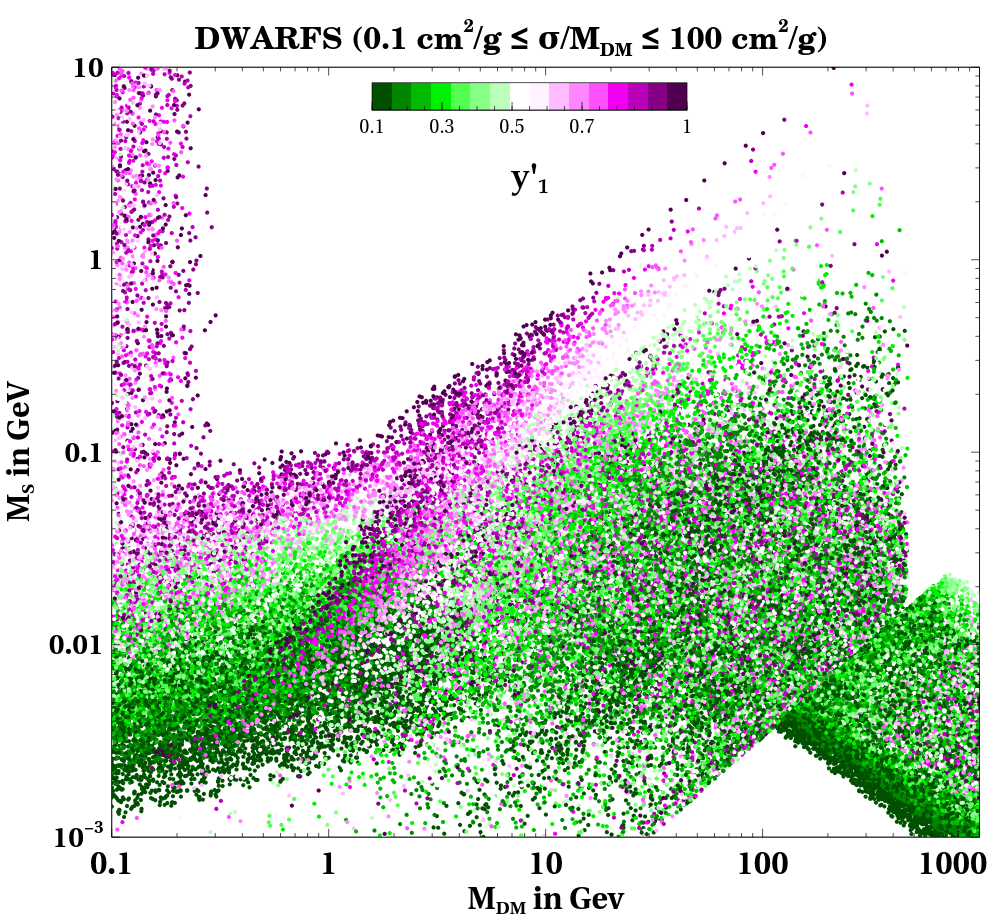}
		%\hfil
		%\includegraphics[scale=0.4]{rhn_sidm_dwarf_rep}
		\caption{ Self-interaction cross-section in the range $0.1-100 \; {\rm cm}^2/{\rm g}$ for dwarfs ($v\sim10 \; {\rm km/s}$). }
		\label{sidm3}
	\end{figure}

	The dark sector particles have elastic self-scattering through t-channel processes due to the presence of the term $y'_1 \overline{N_{R_{1}}}(N_{R_{1}})^c S$ in the model Lagrangian given by Eq.~\eqref{lag}. The Feynman diagram of such process is shown in Fig.~\ref{sidm}.
	In order to alleviate the small-scale anomalies of $\Lambda{\rm CDM}$, the typical DM self-scattering cross-section should be $\sigma \sim 1~{\rm cm}^2 /{\rm g} \approx 2\times 10^{-24}~{\rm cm}^2 /{\rm GeV}$, which is 14 orders of magnitude larger than the typical WIMP cross-section($\sigma \sim 10^{-38} \, {\rm cm}^2/GeV$). This suggests the existence of a light mediator, which is much lighter than electroweak scale. The scalar mediator $S$ in our model serves this purpose. The non-relativistic DM scattering can be well described by the attractive Yukawa potential,
	\begin{equation}
		V(r)= \frac{y'^2_1}{4\pi r}e^{-M_{S}r}
	\end{equation} 
	To capture the relevant physics of forward scattering divergence, the transfer cross-section $\sigma_T$ is defined as~\cite{Feng:2009hw,Tulin:2013teo,Tulin:2017ara}
	\begin{equation}
		\sigma_T = \int d\Omega (1-\cos\theta) \frac{d\sigma}{d\Omega}
	\end{equation}
	Capturing the whole parameter space requires the calculations to be carried out well beyond the perturbative limit. Depending on the masses of DM ($M_{\rm DM}$) and the mediator ($M_{S}$) along with relative velocity of DM (v) and interaction strength ($y'^2_1$), three distinct regimes can be identified, namely the Born regime ($y'^2_1 M_{\rm DM}/(4\pi M_S) \ll 1,  M_{\rm DM} v/M_{S} \geq 1$), classical regime ($y'^2_1 M_{\rm DM}/(4\pi M_S) \geq 1, y'^2_1 M_{\rm DM}/4\pi M_S\geq 1$) and the resonant regime ($y'^2_1 M_{\rm DM}/(4\pi M_S) \geq 1, M_{\rm DM} v/M_{S} \leq 1$). DM self-scattering cross-section in all three regimes are given in Appendix~\ref{appendix2}. Using these self-interaction cross sections and constraining $\sigma/M_{\rm DM}$ in the required range from astrophysical observations at different scales, we get the allowed parameter space of the model for sufficient self-interaction in the plane of DM mass $M_{\rm DM}$ and mediator mass $M_S$. In Fig.~\ref{sidm1} and Fig.~\ref{sidm3}, we show the parameter space for the model in $M_{\rm DM}$ versus $M_S$ plane which gives rise to the required DM self-scattering cross-section ($\sigma/M_{\rm DM}$) in the range $0.1-1~{\rm cm}^2/{\rm g}$ for clusters ($v\sim1000~ \rm km/s$), $0.1-10~{\rm cm}^2/{\rm g}$ for galaxies ($v\sim 200~ \rm km/s$) and $0.1-100~{\rm cm}^2/{\rm g}$ for dwarf galaxies ($v\sim 10~ \rm km/s$). We have also varied the Yukawa coupling $y'_1$ in the range 0.1-1 (shown in coloured bar) which decides the strength of self-scattering.
	
	The allowed region towards the left (right) corner corresponds to the Born(classical) region, where the velocity dependence of the cross-sections are trivial. The central region sandwiched between these two is the resonant region, where quantum mechanical resonances and anti-resonance appear due to the attractive potential. The resonant regime covers a large region of parameter space in the $M_{DM}$ versus $M_S$ plane. These resonances are more prominent at dwarf and galactic scales where DM velocities are comparatively smaller. This is because, for a fixed Yukawa coupling $y'_1$, the condition $M_{DM} v/M_S < 1$ dictates the onset of non-perturbative quantum mechanical effects is easily satisfied by smaller velocities. The resonant spikes are not distinct in these figures as we have varied the Yukawa coupling in a range 0.1-1. Nevertheless, prominent resonant spikes can be seen in Fig.~\ref{sidmdd} in section~\ref{direct}, where we show the same parameter space for a fixed Yukawa coupling $y'_1=0.35$, while confronting the SIDM parameter space to direct search. We can see from the figures that a wide range of DM mass can give rise to sufficient self-interaction. However, the mass of the mediator is constrained roughly within two orders of magnitudes excepting for the resonance case. 
	%We will confront these regions of parameter space to direct detection bounds in section~\ref{sec5}.
	
\begin{figure}
		$$
		\includegraphics[width=8cm,height=6cm]{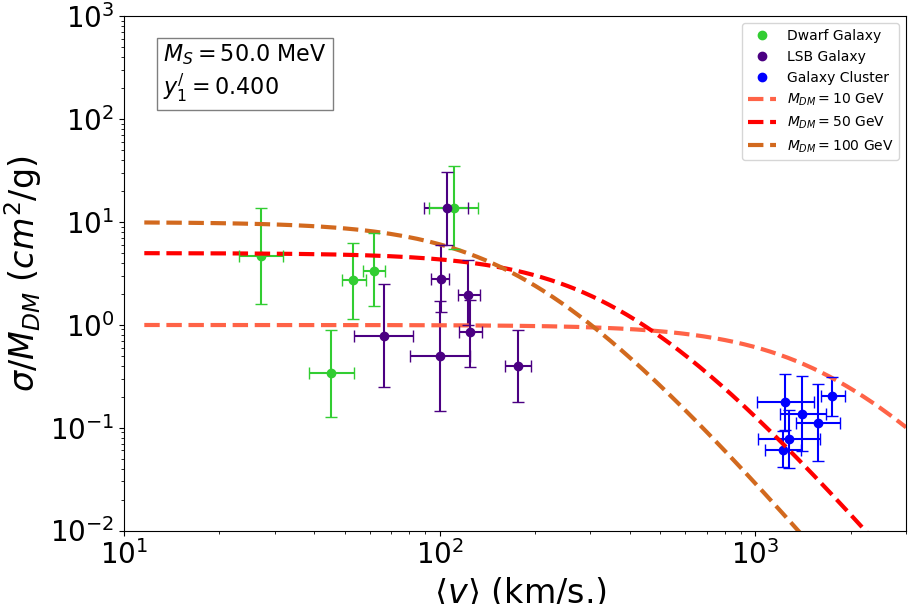}
		$$
		\caption{The self-interaction cross section per unit mass of DM as a function of average collision velocity.}
		\label{astrofit}
	\end{figure}

	The self-scattering cross-section per unit DM mass as a function of average collision velocity is shown in figure~\ref{astrofit}, which fits to data from dwarfs (orange), low surface brightness (LSB) galaxies (blue), and clusters (green)~\cite{Kaplinghat:2015aga,Kamada:2020buc}. The red dashed curve corresponding to the velocity-dependent cross-section calculated from our model for a benchmark point (i.e $M_{\rm DM}=50~\rm GeV$, $M_S=50~\rm MeV$ and $y_1' = 0.4$), which is allowed from all relevant phenomenological constraints, gives a nice fit to the astrophysical observations. It is clear from the Fig.~\ref{astrofit} that the model shows remarkable velocity dependence in self-scattering and can appreciably explain the astrophysical observation of velocity-dependent DM self-interaction.
	
	%%%%%%%%%%%%%%%%%%%%%%%%%%%%%%%%%%%%%%%%%%%%%%%%%%%%%%%%%%%%%%%%%%%%%%%%%%%%%%%%%%%%%%%%%%%%%%%%%%%%%%%%%%%%%%%%%%%%%%%%%%%%%%%%%%%%%%%%%%%%%%%%%%%%%%%%%%%%%%%%%%%%%%%%%%%%%%%%%%%%%%%%%%%%%%%%%%%%%%%%%%%%%%%%%%%%%%%%%%%%%%%%%%%%%%%%%%%%%%%%%%%%%%%%%%%%%%%%%%%%%%%%%%%%%%%%%%%%%%%%%%%%%%%%%%%%%%%%%%%%%%%%%%%%%%%%%%%%%%%%%%%%%%%%%%%%%%%%%%%%%%%%%%%%%%%%%%%%%%%%%%%%%%%%%%%%%%%%%%%%%%%%%%%%%%%%%%%%
	\section{Direct Detection}\label{direct}
\begin{figure}[h!]
		\centering
		\includegraphics[scale=0.25]{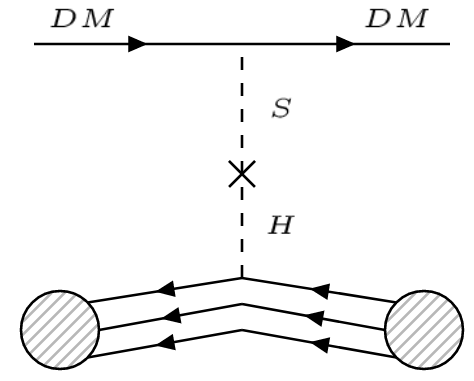}
		\caption{The spin-independent scattering cross-section of DM-nucleon via Higgs portal.}
		\label{DM_diag}
	\end{figure}

	%\label{sec5}
	\label{dd}
	The spin-independent elastic scattering of DM is possible through $S-H$ mixing ( $\theta_{SH}$), where DM particles can scatter off the target nuclei which are located at terrestrial laboratories. The Feynmann diagram for direct detection is shown in Fig.\,\ref{DM_diag} and the scattering cross-section of DM per nucleon can be expressed as
	\begin{equation}
		\sigma_{SI}^{S-H} = \frac{\mu_{r}^{2}}{4\pi A^{2}} \left[ Z f_{p} + (A-Z) f_{n} \right]^{2}
		\label{DD_cs}
	\end{equation}
	
\begin{figure}[h!]
		\centering
		\includegraphics[scale=0.55]{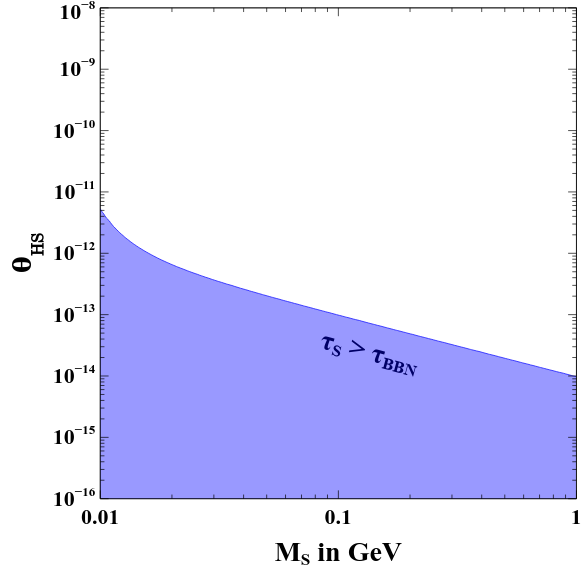}
		\caption{Lifetime of $S$ is shown in the plane of $\theta_{SH}$ versus $M_S$.}
		\label{summary10a}
	\end{figure}

	where $\mu_{r}=\frac{M_{\rm DM}m_n}{M_{\rm DM} + m_n}$ is the reduced mass of the DM-nucleon system, $m_{n}$ being the nucleon (proton or neutron) mass, A and Z are the mass and atomic number of the target nucleus respectively. The $f_{p}$ and $f_{n}$ are the interaction strengths of proton and neutron with DM, respectively and they can be given as,
	\begin{equation}
		f_{p,n}=\sum\limits_{q=u,d,s} f_{T_{q}}^{p,n} \alpha_{q}\frac{m_{p,n}}{m_{q}} + \frac{2}{27} f_{TG}^{p,n}\sum\limits_{q=c,t,b}\alpha_{q} 
		\frac{m_{p,n}}{m_{q}}\,,
		\label{fpn}
	\end{equation}
	where 
	\begin{equation}
		\alpha_{q} =  y'_1 \theta_{SH}\left( \frac{m_{q}}{v}\right) \left[\frac{1}{M^2_S}-\frac{1}{M^{2}_H}\right] \,.
		\label{DD4}
	\end{equation}
	In Eq.\,\eqref{fpn}, the values of $f_{T_{q}}^{p,n}$ can be found in~\cite{Ellis:2000ds} and the mixing angle $\theta_{SH}$ can be derived 
	in terms of the parameters $\lambda_{SH},\langle S \rangle, v, M_S, M_h$. Depending on the value of $\lambda_{SH}$ the $S-H$ mixing can be very small or large. 
	Note that $\theta_{SH}$ also has upper bound by invisible Higgs decay (as the singlet scalar is typically lighter than the Higgs mass), while a lower bound on $\theta_{SH}$ can be obtained by considering $S$ to decay before 
	the big bang nucleosynthesis (BBN) epoch, {\it i.e.} $\tau_S <  \tau_{\rm BBN}$. In Fig. \ref{summary10a}, we have shown the lower bound on $\theta_{SH}$ as a function of $M_S$ from this lifetime criteria. We see from 
	Fig.\ref{summary10a} that $\theta_{SH}<10^{-11}$ is disfavoured for all $M_S \sim 10$ MeV.    
	
	Using Eq.~\eqref{fpn} and \eqref{DD4}, the spin-independent cross-section in Eq.\,\eqref{DD_cs}, can be re-expressed as:
	\begin{eqnarray}
		\sigma_{\rm SI}^{S-H} &=& \frac{{\mu_{r}}^{2} y'^2_1 \theta^2_{SH} }{\pi A^{2}}  \left[ \frac{1}{M^2_{S}}-\frac{1}{M^2_{H}} \right]^{2} \nonumber \\
		& \times &\bigg[ Z \left(\frac{m_{p}}{v}\right) \left(f_{Tu}^{p}+f_{Td}^{p}+f_{Ts}^{p}+\frac{2}{9}f_{TG}^{p} \right) \nonumber\\&+& (A-Z) \left(\frac{m_{n}}{v}\right) \left(f_{Tu}^{n}+f_{Td}^{n}+f_{Ts}^{n}+\frac{2}{9}f_{TG}^{n}\right)  \bigg]^{2}. \nonumber\\
		\label{SI_cross}
	\end{eqnarray}		
	
\begin{figure}[h!]
		\centering
		\includegraphics[width=8cm,height=8cm]{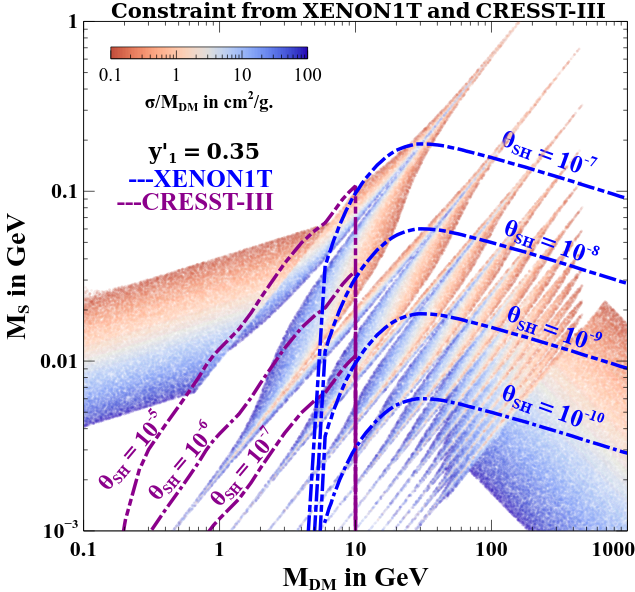}
		%\hfil
		%\includegraphics[scale=0.45]{rhn_sidm_dd_xenon1t_rep}
		\caption{Constraints from DM direct detection in the plane of DM mass $(M_{\rm DM})$ versus mediator mass $(M_S)$ for self-interaction.}
		\label{sidmdd}
	\end{figure}

	Direct search experiments like CRESST-III~\cite{Abdelhameed:2019hmk} and XENON1T \cite{Aprile:2018dbl} put severe constraints on the model parameters. XENON1T provides the most stringent constraint on DM mass above 10 GeV, while CRESST constrains the mass regime below 10 GeV. In Fig.~\ref{sidmdd}. The most stringent constraints from CRESST-III~\cite{Abdelhameed:2019hmk}, XENON1T \cite{Aprile:2018dbl} experiments on $M_{\rm DM}-M_S$ plane are shown against the self-interaction favoured parameter space assuming $y'_1=0.35$. The blue (purple) coloured contours denote exclusion limits from XENON1T (CRESST-III) experiment for specific $S-H$ mixing parameter $\theta_{SH}$. The region below each contour is excluded for that particular $\theta_{SH}$ as shown in the Fig. \ref{sidmdd}. It is seen that direct search experiments severely constrain the allowed parameter space for self-interaction. In particular, for $M_{\rm DM}=50$ GeV and $M_S=50$ MeV, $\theta_{SH} > 10^{-7}$ is already ruled out. 
	
\section{Indirect Detection}
\label{inddet}
Now we briefly discuss about the indirect detection prospects and relevant constraints for the model considered here. Being p-wave process, all the annihilation channels of DM in this model are velocity suppressed ($\sigma v \sim v^2$), unlike a S-wave process, where the cross-section is independent of velocity ($\sigma v \sim v^0$).
Since we are considering the mediator $S$ to be very light for sufficient self-interaction, the annihilation cross-section must be multiplied by the Sommerfeld enhancement factor~\cite{Sommerfeld}, which reflects the modification of the initial-state wave function due to multiple mediator exchange: $(\sigma v)_\text{enh} = S \times \sigma v$. The Sommerfeld enhancement factors for $s$-wave and $p$-wave annihilations are given respectively as~\cite{Cassel:2009wt,Iengo:2009ni,Slatyer:2009vg},
\begin{equation}
	\begin{aligned}
		S_s&=\frac{\pi}{a}\frac{\sinh(2\pi a c)}{\cosh(2\pi ac)-\cos(2\pi\sqrt{c-(ac)^2})}\\
		S_p&=\frac{(c-1)^2+4(ac)^2}{1+4(ac)^2} S_s
	\end{aligned}
\end{equation}
where $a=2\pi v/y'^2$ and $c=3 y'^2 M_{DM}/2\pi^3 M_S$

%for an $s$-wave annihilation process~\cite{Cassel:2009wt,Iengo:2009ni,Slatyer:2009vg} 
%\begin{equation}
% S_s = \frac{\pi}{a} \frac{\sinh (2 \pi \, a \, c)}{\cosh (2 \pi \, a \, c) - \cos (2 \pi \sqrt{c - a^2 c^2})} \; ,
%\end{equation}
%where $a = v/(2 \alpha_\psi)$ and  $c = 6 \, \alpha_\psi \, m_\psi / (\pi^2 m_\phi)$ with $\alpha_\psi = y_\psi^2 \, \cos^2 \delta_\psi / (4\pi)$
%\footnote{Note that we only use the scalar part of the coupling to calculate the Sommerfeld enhancement. We will return to this issue in more detail in the context of DM self-interactions in section~\ref{sec:constraints} and in particular in appendix~\ref{app:pseudoSIDM}.} The corresponding expression for a $p$-wave process is
%\begin{equation}
% S_p = \frac{(c-1)^2 + 4 \, a^2 c^2}{1 + 4 \, a^2 c^2} \times S_s \; . 
%\end{equation}
Note that, for $v \gtrsim \frac{y'^2}{4\pi}$, we get $S_{s,p} \approx 1$ (and hence not significant at the epoch of DM freeze-out), whereas for smaller velocities $S$ increases proportionally to $1/v$ in the $s$-wave case and $1/v^3$ in the $p$-wave case, so that effectively all DM annihilation cross sections (e.g. Eq.~\ref{ann}) increases proportionally to $1/v$ with decreasing velocity in both the cases. The Sommerfeld enhancement saturates for $v \lesssim M_S / (2 M_{\rm DM})$, so the ratio of the two masses determines the maximum possible enhancement. Since all the annihilation channels to SM final states are further suppressed by the scalar mixing $\theta_{SH}$(so that $\langle \sigma v \rangle_{{\rm DM~DM} \to {\rm SM~SM}} \sim \theta^2_{SH}/v$), all fluxes of gamma rays, cosmic rays and neutrinos produced by such a set-up are well below the present and future reach of indirect detection probes~\cite{Arina:2018zcq}, even in presence of maximum Sommerfeld enhancement. For example, we have estimated that, the benchmark values of the parameters ($M_{\rm DM} = 100~ {\rm GeV}, M_S = 10~ {\rm MeV}, y' = 0.35,~ \theta_{SH} \sim 10^{-7}$) considered here gives Sommerfeld enhancement of $\mathcal{O}(10^4)$ and  an effective cross-section of $\mathcal{O}(10^{-40}){\rm cm^3/s}$ for the local galactic DM velocity $v=240~ km/s = 8\times 10^{-4}c$, which is well below the current constraints given by the indirect search experiments like Fermi-LAT~\cite{Fermi-LAT:2015att,Fermi-LAT:2013sme}, MAGIC~\cite{MAGIC:2016xys}, HESS~\cite{HESS:2018cbt}, AMS-02~\cite{2013PhRvL.110n1102A}, constraints from CMB by Planck~\cite{Aghanim:2018eyx} and $ \gamma $-rays by INTEGRAL~\cite{Knodlseder:2007kh}.

	\section{Collider Signatures}
	\label{sec7}
	%%%%%%%%%%%%%%%%%%%%%%%%%%%%%%%%%%%%%%%%%%%%%%%%%%%%
	%%%%%%%%%%%%%%%%%%%%%%%%%%%%%%%%%%%%%%%%%%%%%%%%%%%%
	The fermionic singlet-doublet DM model is rich in collider phenomenology with several interesting signatures such as opposite sign dilepton + missing energy $(\ell^+ \ell^- +\slashed{E_T})$, three leptons + missing energy $(\ell \ell \ell +\slashed{E_T})$ etc~\cite{Dutta:2020xwn, Bhattacharya:2021ltd, Calibbi:2018fqf}. Here we briefly point out an interesting feature of the model: the displaced vertex signature of $\psi^{\pm}$. Note that the doublet mass required to obtain the correct DM relic density (see Fig.~\ref{relic}) is above $\mathcal{O}$(10 TeV). It is not possible to produce such a heavy particle currently at LHC. However, it can produced at proposed HE-LHC~\cite{FCC:2018bvk} with 27 TeV centre-of-mass energy and FCC-hh \cite{FCC:2018vvp} with 100 TeV centre-of-mass energy. Once these particles are produced by virtue of gauge interactions, they can be long-lived before decaying into final state particles including DM \cite{Dutta:2020xwn,Bhattacharya:2017sml,Bhattacharya:2018fus, Borah:2018smz,Calibbi:2018fqf}. The charged component of the doublet, $\psi^{\pm}$ may have a sufficiently long lifetime leading to a displaced vertex signature if produced at colliders. The final states of such displaced vertex in forms of charged leptons or jets can be reconstructed by dedicated analysis, some of which in the context of the Large hadron collider (LHC) may be found in \cite{ATLAS:2016tbt,CMS:2016kce, ATLAS:2016tbt}. Similar analysis in the context of upcoming experiments like MATHUSLA, electron-proton collider and FCC may be found in \cite{Curtin:2017izq,Curtin:2017bxr,Jana:2020qzn,Sen:2021fha} and references therein. 
	
\begin{figure}[ht!]
   \includegraphics[width=8cm]{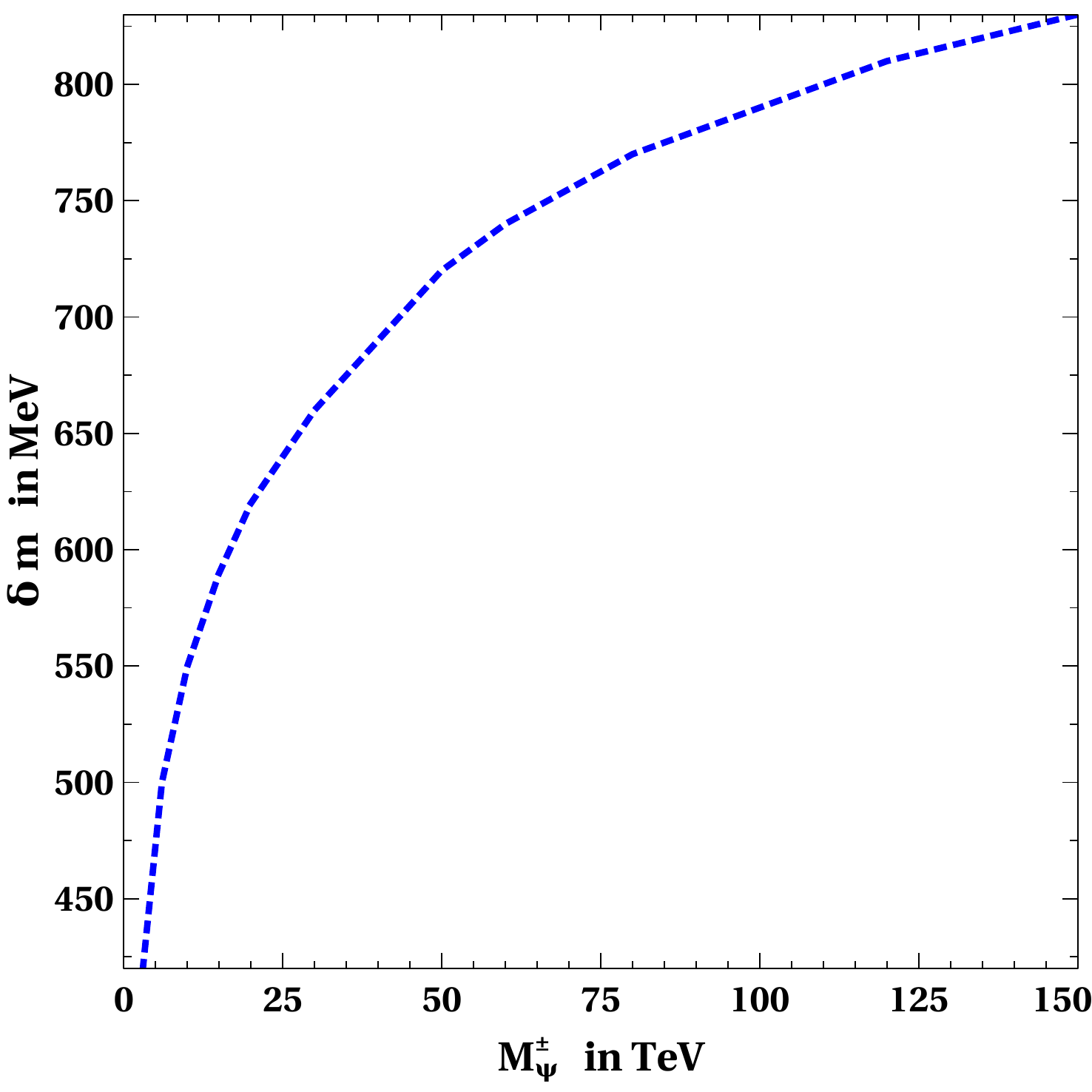}
 \caption{Mass splitting between $\psi^\pm$ and  $\psi^0$ as a function of $M_{\psi^\pm}$.}
 \label{fig:msplit}
\end{figure}

\begin{figure}[h!]
		\centering
		\includegraphics[height=8.5cm,width=8cm]{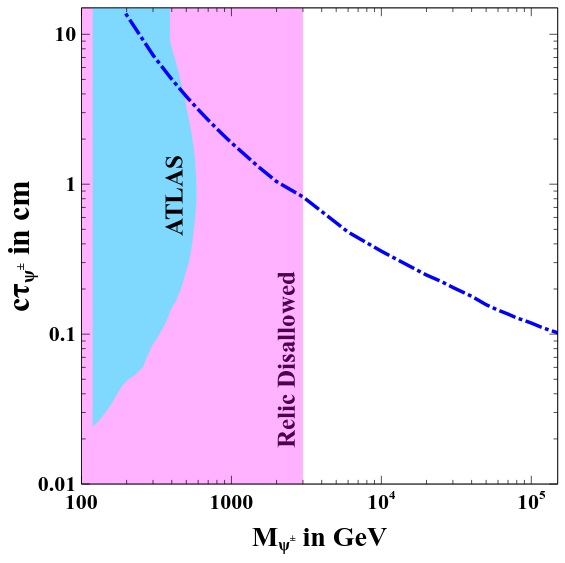}
		\caption{%Top panel: Branching ratio of the decay channel $\psi^\pm \to \pi^\pm \psi^0$, 
		Decay length of $\psi^\pm$ in its rest frame as a function of $M_{\psi^\pm}$}
		\label{DV_log}
	\end{figure}

In the present case, $M_{\psi^0} >> M_W$\footnote{LEP experiment currently excludes charged doublet mass below 102.5 GeV~\cite{Abdallah:2003xe}.}, so the heavy charged states $\psi^\pm$ can decay directly into a $W^\pm$ and the singlets ($\psi^\pm \to W^\pm N_{R_i}$) which is suppressed by the tiny singlet-doublet mixing. Importantly, a mass splitting between the charged component of the doublet $\psi^\pm$ and the neutral component $\psi^0$ can be created from quantum corrections at one loop with virtual photon and Z boson exchange. Virtual $W^\pm$ bosons do not contribute as their couplings to $\psi^\pm$ and $\psi^0$ are identical.
This mass splitting is given by~\cite{Thomas:1998wy,Cirelli:2009uv},
%%%%%%%%%%%%%%%%
\begin{equation}
\delta m = |M_{\psi\pm}| - |M_{\psi^0}| = \frac{\alpha}{2} M_Z f\Big(\frac{M^2_{\psi^\pm}}{M^2_Z} \Big)
\end{equation}
where $f(r)$ is the loop function given as, 
\begin{equation}
f(r)=\frac{r}{\pi}\int^1_0 dx (2-x)~ ln\Bigg[ 1+ \frac{x}{r\sqrt{(1-x)^2}}\Bigg]. \nonumber
\end{equation}
where $\alpha$ is the electromagnetic coupling constant. This mass splitting $\delta m $ is shown as a function of $M_{\psi^\pm}$ in Fig.~\ref{fig:msplit}.
%%%%%%%%%%%%%%%%%%
With this mass splitting, $\psi^\pm$ can decay into the neutral component $\psi^0$ and a soft pion via an off-shell W ($\psi^\pm \to \pi^\pm \psi^0$), which dominates over leptonic decay modes involving $l^\pm \nu $ instead of $\pi^\pm$ ($\psi^\pm \to l^\pm \nu \psi^0$)~\cite{Thomas:1998wy}. The possible decay channels of $\psi^\pm$ and the corresponding decay widths are summarised in Appendix~\ref{appendix4}. We see that, unless the Yukawa coupling ($y_i=y$) is larger than $2\times 10^{-6}$, the decay mode $\psi^\pm \to \pi^\pm \psi^0$ always dominates over other decay channels $\psi^\pm \to W^\pm N_{R_{i}}$ as the latter is suppressed by the singlet-doublet mixing. However larger Yukawa coupling will not give the correct relic density as the doublet would decay much early in that case and would lead to under-abundance. So, for the scale of Yukawa coupling that gives correct relic ($y \sim 10^{-10}$), $\psi^\pm \to \pi^\pm \psi^0$ is the dominant decay mode and its decay width is given by,
\begin{equation}
\Gamma_{\psi^\pm \to \pi^\pm \psi^{0} } =\frac{G^2_F}{\pi}(f_\pi \cos\theta_c)^2  ~\delta m^3 \sqrt{1-\frac{m^2_{\pi^\pm}}{\delta m^2}},\\
\end{equation}  
where, $G_F = 1.16 \times 10^{-5} \; {\rm GeV}^{-2}$ is the Fermi constant, $f_\pi \approx 135$ MeV is the pion form factor, $\theta_c$ is the Cabibbo angle and $m_{\pi^\pm} = 139.57$ MeV is the charged pion mass. 
%For the decay $\psi^\pm \to W^\pm N_{R_i}$, we define $Delta M_i = M_{\psi^\pm} - M_{N_{R_i}}$ (see Appendix \ref{appendix4}). These decay channels start dominating for large Yukawas ($y> 10^{-6}$), $\psi^\pm \to W^{\pm} N_{R_1}$ being the most dominating as $N_{R_1}$ is the lightest one (DM). In calculating the decay width of this channel, we consider $\Delta M$ as allowed by relic density constraint shown in Fig.~\ref{relic}. However for doublet mass range not included in Fig.~\ref{relic} ($M_{\psi^\pm} < 3 TeV$), we have fixed $\Delta M = 10$ GeV.
We show the corresponding decay length $c\tau_{\psi^\pm}$ in the rest frame of $\psi^\pm$ as a function of $M_{\psi^\pm}$ in Fig.~\ref{DV_log}. The decay length varies within 0.1-10 cm which gives rise to a displaced vertex signature at colliders. The cyan coloured region is disallowed from ATLAS search for such long-lived charged particles with a lifetime ranging from 10 ps to 10 ns~\cite{Aaboud:2017mpt}. As shown in Fig.~\ref{relic}, the doublet mass below 3 TeV can not give rise to correct DM relic density. We exclude these $\psi^\pm$ mass range as depicted by the magenta coloured region. Such displaced vertex signatures can be probed as a signature of verifiability of the model at present and future colliders. 
	
	\section{Conclusion}\label{sec8}
	%%%%%%%%%%%%%%%%%%%%%%%%%%%%%%%%%%%%%%%%
	We have studied a singlet-doublet fermion dark matter model to explain the self-interacting nature of dark matter and sub-eV masses of light neutrinos simultaneously. We extended the SM with a vector-like fermion doublet and three right-handed neutrinos (RHNs), all odd under an imposed $\mathcal{Z}_2$ symmetry. We assumed a negligible mixing between the fermion singlet and fermion doublet in order to keep the doublet long-lived. Moreover, the singlet RHNs are much lighter than the doublet so that the lightest RHN serves as a candidate of DM. Light scalar $S$ ($M_S < M_{\rm DM})$ having sizeable Yukawa coupling ($\sim \mathcal{O}(0.1)$ with the DM not only facilitates velocity-dependent DM self-interaction that helps alleviating the small-scale anomalies of $\Lambda {\rm CDM}$, but also mixes with the SM Higgs providing a portal for detecting such SIDM at terrestrial direct search laboratories. We show that for a typical SIDM of mass 50 GeV and a mediator mass 50 MeV, direct detection experiments like XENON1T and CRESST-III already rule out scalar portal mixing $\theta_{SH} > 10^{-7}$. Due to the large coupling of DM with the mediator, the thermal relic of the dark matter is negligibly small as DM annihilates efficiently into the light mediator. However, due to the small mixing between the singlet and doublet fermions, the thermal relic of long-lived fermion doublet gets converted into singlet DM at late epochs, typically before the BBN but after the thermal freeze-out of SIDM. The doublet is required to be very heavy in order to generate the correct DM relic density and can be pair- produced at future collider experiments such as HE-LHC and FCC-hh, the decay of which may give rise to displace vertices.All annihilation channels being p-wave suppressed, the model is also safe from bounds by the indirect detection experiments even in presence of Sommerfeld enhancement due to multiple mediator exchanges. While the lightest RHN is the SIDM, three copies of RHNs along with a $\mathcal{Z}_2$-odd scalar doublet can lead to the generation of light neutrino mass at a one-loop level. These $\mathcal{Z}_2$-odd particles, apart from their typical contributions to charged lepton flavour violating signatures, can also play an interesting role in DM relic evolution as discussed in this work.

\vspace*{0.8cm}	
	
	\acknowledgements
	%%%%%%%%%%%%%%%%%%%%%%%%%%%%%%%%%%%%%%%The 
	MD acknowledges Department of Science and Technology (DST), Govt. of India, for providing the financial assistance for the research under the grant DST/INSPIRE/03/ 2017/000032. NS acknowledges the support from Department of Atomic Energy (DAE)-Board of Research in Nuclear Sciences (BRNS), Government of 
India (Ref. Number: 58/14/15/2021-BRNS/37220).
	
	%\newpage
	
	\appendix
	%\begin{widetext}
	%\section{Relevant cross section and decay widths}

	\section{Interactions for DM relic calculations}
	\label{appendix1}
	Gauge interaction of $\Psi$ with SM is given by:
	
	\begin{equation}
		\begin{aligned}
			\mathcal{L}^\Psi_{int} &=\overline{\Psi}i\gamma^\mu(-i\frac{g}{2}\tau.W_\mu - ig'\frac{Y}{2}B_\mu )\Psi
			\\ 
			&= \Big(\frac{e}{2\sin\theta_W \cos\theta_W}\Big)\overline{\psi^0}\gamma^\mu Z_\mu \psi^0
			\\
			& +\frac{e}{\sqrt{2}\sin\theta_W}(\overline{\psi^0}\gamma^\mu W^+_\mu \psi^- + \psi^+\gamma^\mu W^-_\mu \psi^0) \\
			& - e \psi^+\gamma^\mu A_\mu \psi^-\\ 
			& -\Big(\frac{e \cos2\theta_W}{2\sin\theta_W \cos\theta_W}\Big)~ \psi^+\gamma^\mu Z_\mu \psi^- .
		\end{aligned}
	\end{equation}
	where $g=\frac{e}{\sin\theta_W}$ and $g'=\frac{e}{\cos\theta_W}$, $e$ being the electromagnetic coupling constant and $\theta_W$, the Weinberg angle. 
	
	\section{DM Self-interaction cross sections at low energy}
	\label{appendix2}
	In the Born Limit ($y'^2_1 M_{DM}/(4\pi M_S) << 1$),
	\begin{equation}
		\sigma^{Born}_T=\frac{y'^4_1}{2\pi M^2_{DM} v^4}\Bigg(ln(1+\frac{ M^2_{DM} v^2}{M^2_S})-\frac{M^2_{DM}v^2}{M^2_S+ M^2_{DM}v^2}\Bigg)
	\end{equation} 
	Outside the Born regime ($y'^2_1 M_{DM}/(4\pi M_S) \geq 1 $), there are two distinct regions {\it viz}, the classical regime and the resonance regime. In the classical regime ($y'^2_1 M_{DM}/4\pi M_S\geq 1, M_{DM} v/M_{S} \geq 1$), the solutions for an attractive potential is given by\cite{Tulin:2013teo,Tulin:2012wi,Khrapak:2003kjw}:
	\begin{equation}
		\sigma^{classical}_T =\left\{
		\begin{array}{l}
			
			\frac{4\pi}{M^2_S}\beta^2 ln(1+\beta^{-1}) ~~~~~~~~~~~\beta \leqslant 10^{-1}\\
			\frac{8\pi}{M^2_S}\beta^2/(1+1.5\beta^{1.65}) ~~~~~~~~10^{-1}\leq \beta \leqslant 10^{3}\\
			\frac{\pi}{M^2_S}( ln\beta + 1 -\frac{1}{2}ln^{-1}\beta) ~~~~~\beta \geq 10^{3}\\
		\end{array}
		\right.
	\end{equation}  
	
	where $\beta = 2y^2_1 M_{N_R}/(4\pi M_S) v^2$.
	
	%Both the Born and the classical regime do not provide us the mild velocity dependence in the cross-section required to explain the small scale issues. However one interesting 
	
	Outside the classical regime, there lies the resonant regime for ($y'^2_1 M_{DM}/(4\pi M_S) \geq 1, M_{DM} v/M_{S} \leq 1$), characterised by the appearance of quantum mechanical resonances and anti-resonance in $\sigma_T$ due to (quasi-)bound states formation in the attractive potential. No analytical formula for $\sigma_T$ is available in this regime and the non-relativistic Schrodinger equation needs to be solved by partial wave analysis. Instead, here we use the non-perturbative results obtained by approximating the Yukawa potential to be a Hulthen potential $\Big( V(r) = \pm \frac{y'^2_1}{4\pi}\frac{ \delta e^{-\delta r}}{1-e^{-\delta r}}\Big)$, which is given by~\cite{Tulin:2013teo}:
	\begin{equation}
		\sigma^{Hulthen}_T = \frac{16 \pi \sin^2\delta_0}{M^2_{DM} v^2}
	\end{equation}
	where $l=0$ phase shift $\delta_0$ is given in terms of the $\Gamma$ functions by :
	\begin{eqnarray}
		\delta_0 &=arg \Bigg(i\Gamma \Big(\frac{i M_{DM} v}{k~ M_S}\Big)\bigg/{\Gamma (\lambda_+)\Gamma (\lambda_-)}\Bigg)\nonumber\\
		\lambda_{\pm} &=
		\begin{array}{l}
			1+ \frac{i M_{DM} v}{2 ~k ~M_S}  \pm \sqrt{\frac{\alpha_D M_{DM}}{k M_S} - \frac{ M^2_{DM} v^2}{4 k^2 M^2_S}}\\
		\end{array}
	\end{eqnarray}   
	and $k \approx 1.6$ is a dimensionless number. 
	
	%\begin{widetext}
	\onecolumngrid
	\section{Relevant cross-sections and decay widths for relic density calculation}
	\label{appendix3}
	\begin{eqnarray}
		\Gamma({\Psi \rightarrow H N_i}) &=& \frac{y^2_i}{32\pi M^3_{\Psi}}\big((M_{\Psi} + M_{N_i})^2 - M^2_H \big)
		\\\nonumber&\times&\big( M^4_{\Psi} + M^4_{N_i} + M^4_H - 2 M^2_{\Psi} M^2_{N_i} - 2 M^2_{N_i} M^2_H - 2 M^2_{\psi} M^2_H \big)^{\frac{1}{2}}\\
		\Gamma({\eta \rightarrow N_i l}) &\approx& \frac{Y^2_{\alpha i}}{32 \pi}M_{\eta}\Big(1-4\frac{M^2_{N_i}}{M^2_{\eta}} \Big)^{3/2}\\
		\Gamma({N_{2,3} \rightarrow N_1 l~l}) &\approx& \frac{Y^2_{\alpha 1} Y^2_{\alpha 2,3}}{8 \pi}\frac{(\Delta M_{2,3})^5}{M^4_\eta}~~~~~ ({\rm ~~where} ~ \Delta M_{2,3}=M_{N_{2,3}}-M_{N_1}~~).
	\end{eqnarray}
	\begin{eqnarray}	
		\sigma({ N_i \;N_i} \rightarrow S S) &=& \frac{y'^4_i}{192 \pi s (s-4M^2_{N_i})} \times \Bigg[\frac{24s(4M^4_{N_i}+2M^4_{S}+sM^2_{N_i})A}{M^4_{S}+M^2_{N_i}s-4M^2_{S}M^2_{N_i}}\nonumber\\ & -&\frac{24 (8M^2_{N_i}-4M^2_{S}-s^2-(s-2M^2_{S})4M^2_{N_i})}{s-2M^2_{S}} {\rm Log}\Big[\frac{2M^2_{S}+s(A-1)}{2M^2_{S}-s(A+1)}\Big]\Bigg]\nonumber \\
	\end{eqnarray}
	where $A=\sqrt{\frac{(s-4M^2_{S})(s-4M^2_{N_i})}{s^2}}$.	
	Thermal averaged cross-section for annihilation of $A$ to $B$  is given by: \cite{Gondolo:1990dk}
	\begin{equation}
		\langle\sigma v \rangle_{AA \rightarrow BB} = \frac{x}{2\big[K^2_1(x)+K^2_2(x)\big]}\times \int^{\infty}_{2}  dz \sigma_{(AA\rightarrow BB)} (z^2 - 4)z^2 K_1(zx)% (z^2 M^2_A)
		\label{appeneq1}
	\end{equation}
	where $z=\sqrt{s}/M_A$ and $x=M_A/T$.	
	Thermal averaged decay width of $A$ decaying to $B C$ is given by:
	\begin{equation}
		\langle \Gamma(A \rightarrow B C) \rangle = \Gamma(A \rightarrow B C)  \Bigg(\frac{K_1(x)}{K_2(x)}\Bigg)
		\label{appeneq2}
	\end{equation}	
	In Eqn. \eqref{appeneq1} and \eqref{appeneq2}, $K_1$ and $K_2$ are the modified Bessel functions of 1st and 2nd kind respectively.
	
	%\end{widetext}
\section{Possible decay modes of $\psi^\pm$}
	\label{appendix4}	%\twocolumngrid
	
The decay widths of possible decay modes of $\psi^\pm$ are \cite{Thomas:1998wy,Calibbi:2018fqf},

	\begin{eqnarray}
		%\begin{aligned}
		\Gamma_{\psi^\pm \to \pi^\pm \psi^{0} } &=&\frac{G^2_F}{\pi}(f_\pi \cos\theta_c)^2  ~\delta m^3 \sqrt{1-\frac{m^2_{\pi^\pm}}{\delta m^2}},\\
			\Gamma_{\psi^{\pm} \to \psi^0 l^{\pm} \nu} &= &
        \frac{G_F^2}{15\pi^3}\delta m^5
         \sqrt{1-b^2_l} P(b_l),\\
			\Gamma_{\psi^\pm \to \pi^{\pm} N_{R_i}}  &\approx & 2y^2_i v^2 \frac{G^2_F}{\pi}(f_\pi \cos\theta_c)^2   ~\Delta M_i \sqrt{1-\frac{m^2_{\pi^\pm}}{\Delta M^2_i}},\\
			\Gamma_{\psi^\pm \to W^{\pm} N_{R_i}}  &\approx & \frac{\alpha y^2_i v^2}{32 s^2_{W}} ~\sqrt{\lambda(M^2_{\psi^\pm},M_{N_{R_i}}^2,M^2_W)}
\frac{\left((M_{\psi^\pm} + M^2_{N_{R_i}} + 2 M^2_W \right) \left((M_{\psi^\pm} - M_{N_{R_i}})^2-M^2_W\right)}{M^3_{\psi^\pm} M^2_W (M_{\psi^0}-M_{N_{R_i}})^2}
		%\end{aligned}
		\label{decay}
	\end{eqnarray}
where ,
$P(b_l)= 1-\frac{9}{2}b_l^2-4b_l^4+   \frac{15b_l^4}{2\sqrt{1-b_l^2}} {\rm tanh}^{-1}\sqrt{1-b_l^2}$, $b_l =m_l/\delta m$,\\ $\lambda (a,b,c)= a^2+b^2+c^2-2 ab -2 ac -2 bc$, $G_F = 1.16 \times 10^{-5} \; {\rm GeV}^{-2}$ is the Fermi constant, $f_\pi \approx 135$ MeV is the pion form factor, $\alpha$ is electromagnetic coupling constant, $\theta_c$ is the Cabibbo angle, $v=246 {\rm GeV}$ is the vacuum expectation value (vev), $s_W$ is the sin of Weinberg angle, $\delta m = M_{\psi\pm} - M_{\psi^0}$, $\Delta M_i =  M_{\psi^\pm} - M_{N_{R_i}} $, $M_W$ is the mass of W boson and $m_{\pi^\pm} = 139.57$ MeV is the charged pion mass.

\section{Details of CLFV Decay $\boldsymbol{\mu \to 3e}$}
\label{mu23e}
\begin{eqnarray}
	\text{Br}\left(\mu \to
	e \overline{e}e\right)&=&
	\frac{3(4\pi)^2\alpha^2}{8G_F^2} ~M^2~
	\mathrm{Br}\left(\mu \to e\nu_{\mu}
	\overline{\nu_e}\right) \\
\end{eqnarray}
where $M^2$ is given as:
\begin{eqnarray}
M^2&=&\left[|A_{ND}|^2
+|A_D|^2\left(\frac{16}{3}\log\left(\frac{m_\mu}{m_e}\right)
-\frac{22}{3}\right)+\frac{1}{6}|B|^2\right.\nonumber\\
&&\left.+ \frac{1}{3} \left( 2  |F_{RR}|^2 + |F_{RL}|^2 \right)
+\left(-2 A_{ND} A_D^{*}+\frac{1}{3}A_{ND} B^*
-\frac{2}{3}A_D B^*+\mathrm{h.c.}\right)\right]
\end{eqnarray}
where $A_D$ is as given in Eq.~\ref{ADMEG}, and $A_{ND}$ is given by:
\begin{equation}
	A_{ND}=\sum_{k=1}^3\frac{(Y)^*_{ke} (Y)_{k\mu}}
	{6(4\pi)^2}\frac{1}{M_{\eta^+}^2}
	G_2\left(r_k\right), \label{eq:A1L}
\end{equation}
and 
\begin{equation}
	F_{RR} = \frac{F \, g_R^\ell}{g_2^2 \sin^2 \theta_W M_Z^2} \qquad ,
	\qquad F_{RL} = \frac{F \, g_L^\ell}{g_2^2 \sin^2 \theta_W M_Z^2} \quad,
\end{equation}
with the co-efficient $F$ given by
\begin{equation}
	F = \sum_{k=1}^3\frac{(Y)^*_{ke} (Y)_{k\mu}}
	{2(4\pi)^2}\frac{m_\mu m_e}{M_{\eta^+}^2} \frac{g_2}{\cos \theta_W}
	F_2\left(r_k\right) \, .
	\label{eq:FR}
\end{equation}

For the Box diagrams, the co-efficient $B$ is given by:
{\small \begin{align}
		B = \frac{1}{(4\pi)^2 e^2 M_{\eta^+}^2} 
		\sum_{j,\:k=1}^3\left[\frac{1}{2} D_1(r_j,r_k) (Y)_{k e}^* (Y)_{k e}
		(Y)_{j e}^* (Y)_{j \mu} + \sqrt{r_j r_k}
		D_2(r_j,r_k) (Y)_{k e}^* (Y)_{k e}^* (Y)_{je}
		(Y)_{j \mu}\right].
\end{align}}	
The loop functions $G_2, F_2, D_1, D_2$ are given by:
\begin{eqnarray}
	F_2(x) &=& \frac{1-6x+3x^2+2x^3-6x^2 \log x}{6(1-x)^4}, \\
	G_2(x) &=& \frac{2-9x+18x^2-11x^3+6x^3 \log x}{6(1-x)^4}, \\
	D_1(x,y) &=& - \frac{1}{(1-x)(1-y)} - \frac{x^2 \log x}{(1-x)^2(x-y)} -
	\frac{y^2 \log y}{(1-y)^2(y-x)}, \\
	D_2(x,y) &=& - \frac{1}{(1-x)(1-y)} - \frac{x \log x}{(1-x)^2(x-y)} -
	\frac{y \log y}{(1-y)^2(y-x)}.
\end{eqnarray}

\vspace{0.25cm}

These loop functions do not have any poles. In the limit $x,y\to1$ and
$y\to x$, the functions become 
\begin{eqnarray}
	F_2(1)&=&\frac{1}{12},\quad
	G_2(1)=\frac{1}{4},\quad
	D_1(1,1)=-\frac{1}{3},\quad
	D_2(1,1)=\frac{1}{6},
\end{eqnarray}
\begin{eqnarray}
	&&D_1(x,x)=\frac{-1+x^2-2x\log{x}}{(1-x)^3},~~~
	D_1(x,1)=D_1(1,x)=\frac{-1+4x-3x^2+2x^2 \log{x}}{2(1-x)^3},\\
	&&D_2(x,x)=\frac{-2+2x-(1+x)\log{x}}{(1-x)^3},~~~D_2(x,1)=D_2(1,x)=\frac{1-x^2+2x\log{x}}{2(1-x)^3}.
\end{eqnarray}

\section{Details of $\pmb{\mu \to e}$ conversion in nuclei}
\label{mutoe}

	\begin{align}
		{\rm CR} (\mu \to e, {\rm Nucleus}) &= 
		\frac{p_e \, E_e \, m_\mu^3 \, G_F^2 \, \alpha_{\mathrm{em}}^3 
			\, Z_{\rm eff}^4 \, F_p^2}{8 \, \pi^2 \, Z} K^2 
		\frac{1}{\Gamma_{\rm capt}}\,.
	\end{align}
where	
	\begin{align}
	K^2 &= 
	 \left| (Z + N) \left( g_{LV}^{(0)} + g_{LS}^{(0)} \right) + 
	(Z - N) \left( g_{LV}^{(1)} + g_{LS}^{(1)} \right) \right|^2 + 
 \left| (Z + N) \left( g_{RV}^{(0)} + g_{RS}^{(0)} \right) + 
	(Z - N) \left( g_{RV}^{(1)} + g_{RS}^{(1)} \right) \right|^2 
\end{align}

\vspace{0.5cm}

In the above, $g_{XK}^{(0)}$ and $g_{XK}^{(1)}$ (with $X = L, R$ and $K = S, V$) are given by
\begin{align}
	g_{XK}^{(0)} = \frac{1}{2} \sum_{q = u,d,s} \left( g_{XK(q)} G_K^{(q,p)} +
	g_{XK(q)} G_K^{(q,n)} \right)\,,~~~~ 
	g_{XK}^{(1)} = \frac{1}{2} \sum_{q = u,d,s} \left( g_{XK(q)} G_K^{(q,p)} - 
	g_{XK(q)} G_K^{(q,n)} \right)\,.
\end{align}

\vspace{0.5cm}

Neglecting the Higgs-penguin contributions due to the smallness of the
involved Yukawa couplings. Therefore, the corresponding couplings are
\begin{eqnarray}
	g_{LV(q)} = g_{LV(q)}^{\gamma} + g_{LV(q)}^{Z}\,, ~~
	g_{RV(q)} = \left. g_{LV(q)} \right|_{L \leftrightarrow R}\,,~~ 
	g_{LS(q)} \approx 0 \, , ~~
	g_{RS(q)} \approx 0 \, .
\end{eqnarray}

\vspace{0.25cm}

The photon and $Z$-boson couplings can be computed from the Feynman
diagrams which are given by:
\begin{align}
	g_{LV(q)}^{\gamma} = \frac{\sqrt{2}}{G_F} e^2 Q_q 
	\left(A_{ND} - A_D \right)\,,~~~~
	g_{RV(q)}^{Z} = -\frac{\sqrt{2}}{G_F} \, \frac{g_L^q + g_R^q}{2} \, 
	\frac{F}{M_Z^2} \,. 
\end{align}

%\vspace{0.01cm}

And the tree-level $Z$-boson couplings to a pair of quarks are:
\begin{equation}
	g_L^q = \frac{g_2}{\cos \theta_W}\left( Q_q \sin^2 \theta_W - T_3^q
	\right), \qquad g_R^q = \frac{g_2}{\cos \theta_W} Q_q \sin^2 \theta_W,
\end{equation}

%	\vspace{2cm}
\twocolumngrid

\newpage
	
%	\bibliographystyle{apsrev}
%	\bibstyle{apsrev}
%	\bibliography{ref,ref1,ref12,ref2,ref3,ref4,ref5}
%	\end{document}

\end{document}